\documentclass[11pt,a4paper]{article}
\usepackage{jcappub}
\usepackage{epstopdf}
\usepackage{graphicx}
\usepackage{bm}
\usepackage{epstopdf}
\usepackage{mathtools}
\usepackage{natbib}
\usepackage{captcont}
\usepackage{booktabs}
\usepackage{threeparttable}
\title{Constraints on the Sommerfeld- enhanced dark matter annihilation from the gamma rays of subhalos and dwarf galaxies}
\author[a]{Bo-Qiang Lu,}
\author[a,b]{Yue-Liang Wu,}
\author[a,b]{Wei-Hong Zhang}
\author[a,b]{and Yu-Feng Zhou}
\affiliation[a]{State Key Laboratory of Theoretical Physics, 
Institute of Theoretical Physics, Chinese Academy of Sciences Beijing, 100190, P.R. China}
\affiliation[b]{University of Chinese Academy of Sciences, Beijing, 100049, P.R. China}

\emailAdd{bqlu@itp.ac.cn}
\emailAdd{ylwu@itp.ac.cn}
\emailAdd{whzhang@itp.ac.cn}
\emailAdd{yfzhou@itp.ac.cn}

\abstract{
The substructures of the Galactic dark matter halo such as dark matter subhalos and dwarf galaxies have very low velocity dispersions,
which makes them useful in constraining the scenario of Sommerfeld-enhanced dark matter annihilation.
We calculate the velocity distribution of dark matter particles in the dark matter halo substructures using the Eddington's formula with NFW density profile.
We parameterize the effect of Sommerfeld enhancement of $s$-wave dark matter annihilation on the gamma-ray flux as the Sommerfeld-enhanced $J$-factors,
and explicitly calculate their values for 15 known dwarf spheroidal galaxies.
Using the results from the analysis of the unassociated point-sources in the Fermi-LAT 3FGL and Fermi-LAT gamma-ray data observation on the 15 dwarf galaxies, we derive upper limits on the dark matter annihilation cross sections with Sommerfeld enhancement.
We find that in a wide region of parameter space, the constraints can be a few orders of magnitude more stringent than that in the case without the Sommerfeld enhancement.
For dark matter annihilation dominantly into $e^+e^-$, $\mu^+\mu^-$, $\tau^{+}\tau^{-}$ and $b\bar b$, those constraints can exclude thermal relic dark matter for the dark matter mass below about 1 TeV.
}

\keywords{dark matter, gamma-ray, Sommerfeld enhancement, velocity distribution}

\arxivnumber{}

\begin{document}
\maketitle

\section{Introduction}

Cosmological and astrophysical observations have shown that about 85$\%$ of the matter in the Universe consists of cold dark matter (DM) rather than Standard Model (SM) particles \cite{Jungman1996, Bergstrom2000, Bertone2005RP}. The most popular class of dark matter particle candidates is that of weakly interacting massive particles (WIMPs) \cite{Lee1977, Hut1977}. These particles decouple from the thermal bath as the early Universe is expanding and cooling and finally achieve the appropriate relic density. In this scenario, the observed dark matter abundance determines the dark matter annihilation cross section, provided that the dark matter particles are massive enough to become non-relativistic at freeze-out. The particles such as high energy gamma rays, pairs of electrons and positrons and antiprotons produced in the annihilation of dark matter in regions of high dark matter density, including the Galactic center 
\cite{Merritt2002, Cesarini2004, Dodelson2008}, subhalos in the Milky Way \cite{Stoehr2003, Bergstrom1998}, and dark matter mini-spikes around intermediate-mass black holes \cite{Bertone2005PRD, Brun2007, Bringmann2009} provide a valuable chance to study the nature of dark matter indirectly through accurate measurements of cosmic-rays spectra \cite{Bringmann2012, Klasen2015}.

In the early Universe, small matter density perturbations grow via gravitational instability giving rise to cold, collisionless dark matter bound structures known as halos. N-body simulations under the standard $\Lambda$CDM cosmological framework demonstrate that dark matter halos form hierarchically \cite{Frenk2012}, with low-mass halos forming first and then large-mass halos resulting from the merging and accretion of those smaller halos. The hierarchical nature of the structure formation implies that halos contain very large numbers of smaller subhalos, which orbit within the potential well of a more massive host halo \cite{White1991, Frenk2012}. Subhalos can play a crucial role in indirect dark matter searches, since the cosmic-rays fluxes produced from the self-annihilation of dark matter are proportional to the square of the dark matter density and hence, the presence of substructure could lead to an enhancement over the expected signal from the smooth distribution of dark matter in the host halo \cite{Moline2017}.

Under the assumption that the dark matter annihilation cross section (the cross section multiply by the velocity) is velocity-independent, the gamma-ray flux from the annihilation of dark matter in a subhalo can be expressed as a product of the line-of-sight integral to the dark matter distribution of a subhalo, i.e., the $J$-factor and a component depending on the particle physics models on dark matter annihilation \cite{Ackermann2012}. In this case, the $J$-factor is independent of the underlying dark matter particle physics such as the dark matter mass and cross section, thus, the $J$-factor can be simply factorized from the particle physics.
However, in generic cases the dark matter cross section can be velocity-dependent, for instance, in some models the dark matter annihilation cross section is $p$-wave suppressed (i.e., $ \sigma v\propto v^2$) \cite{Diamanti2014, Boehm2004-2, Robertson2009, Alves2014, Berlin2014-1}. Furthermore, it has been shown that the dark matter annihilation cross section may be enhanced at low relative velocities by the so-called Sommerfeld enhancement, which results from the exchange of light mediators between dark matter particles \cite{Sommerfeld1931, Hisano2004, Hisano2005, Profumo2005, Cirelli2007, Hamed2009, Feng2010-1, Feng2010-2, Zavala2010, Zhou2013PRD, Zhou2013JCAP, Das2017}. 
The Sommerfeld enhancement provides a physical mechanism for the dark matter explanation of the rising positron fraction at energies $\gtrsim 10\rm\;GeV$ observed by the PAMELA and AMS-02 experiments \cite{PAMELA2009, AMS2013, Bergstrom2009, Ibarra2014, Mauroa2014, Jin2013, Yuan2013}.
When the annihilation is velocity-dependent, the produced cosmic-ray flux is also affected by the distribution of dark matter particle velocities, which depends on the location in the subhalo, thus the dark matter annihilation cross section cannot be extracted from the $J$-factor directly \cite{Ferrer2013, Boddy2017}.

In this work we focus on the dark matter annihilation cross section that is enhanced by the Sommerfeld effect. We determine the so-called Sommerfeld-enhanced $J$-factor and put constraints on the particle parameter space using the null results from the Fermi-LAT experiment observing the gamma-ray sources.
To this end, in section 2 we firstly make a brief review of the Eddington's formula. Using this formula, we can determine the dark matter velocity distribution of a subhalo for a given dark matter density profile, with the assumption that the orbits of dark matter particles are isotropic. We define a dimensionless dark matter velocity distribution function and show that this distribution function can be well fitted by an exponential form of velocity.
In section 3, we display our study on the Sommerfeld enhancement and the Sommerfeld-enhanced $J$-factor is presented in section 4.
Finally, we take advantage of the Fermi-LAT dark matter searches of gamma-ray sources to put limits on the Sommerfeld enhancement in section 5. 
For the subhalo observations, we count the numbers of sources that may be observed by the Fermi-LAT experiment and use this to determine the 95\% confidence level Poisson upper limit on the predicted numbers of such sources. For the dwarf satellite galaxies searches, we use the likelihood and upper limits on the gamma-ray flux provided by the Fermi collaboration to determine the upper limits on the dark matter parameters space at 95\% confidence level. With these results, we show that the Sommerfeld enhancement parameter spaces that may account for the positron anomaly have been excluded by the Fermi-LAT gamma-ray observation results. In section 6, we summarize our conclusions.

%*************************************************************************

\section{Dark matter velocity distribution function}

\subsection{Eddington's formula}
For the purpose of determination of the velocity distribution of dark matter, we assume that the gravitational potential that bounds the dark matter particles is spherically-symmetric and the orbits of dark matter particles are isotropic. 
Simulation studies of the velocity anisotropy profiles of the dark matter subhalos are consistent with this range out to the subhalo virial radius \cite{Campbell2017}.
With these assumptions, the particle velocity distribution function can be uniquely determined using the Eddington's formula \cite{Binney2008} for a given mass density profile $\rho(r)$
\begin{eqnarray}
f(\varepsilon )=\frac{1}{\sqrt{8}\pi^2}\int_{\varepsilon }^{0}\frac{d^2\rho}{d\Psi^2}\frac{d\Psi}{\sqrt{\Psi-\varepsilon }}.
\end{eqnarray}
where $\Psi(r)$ represents the spherically-symmetric gravitational potential at position $r$, $f(\varepsilon)\equiv f(r,v)$ is the dark matter velocity distribution function, $v$ is the dark matter particle velocity (in unit c) and $\varepsilon=v^2/2+\Psi(r)<0$ is the gravitational binding energy per mass of a dark matter particle.
The dark matter density profile and velocity distribution function satisfy the normalization $\rho(r)=\int f(r,v)d^3\vec{v}$.
For an isotropic velocity distribution function $f(r,v)$, the normalization can be written as
\begin{eqnarray}
\rho(r)=4\pi\int_{0}^{v_{\rm esc}} v^2f(r,v)dv,
\end{eqnarray}
where $v_{\rm esc}=\sqrt{-2\Psi(r)}$ is the escape velocity of a dark matter particle at radius $r$ in the gravitational binding system.

In the following, the Navarro-Frenk-White (NFW) profile \cite{NFW1997} is used to approximate the dark matter density distribution. For such a distribution of dark matter, $f(r,v)$ can be expressed as a product of a dimensionless distribution and a scale factor. We also find that the dimensionless velocity distribution can be parameterized by an exponential form of velocity. In this work, the parameterization of the dimensionless velocity distribution (i.e. Eq. (A.7), see appendix A for more details) is adopted for the calculation of $f(r,v)$.

%*****************************fig1***************************************
\begin{figure}
\centering
\includegraphics[width=80mm,angle=0]{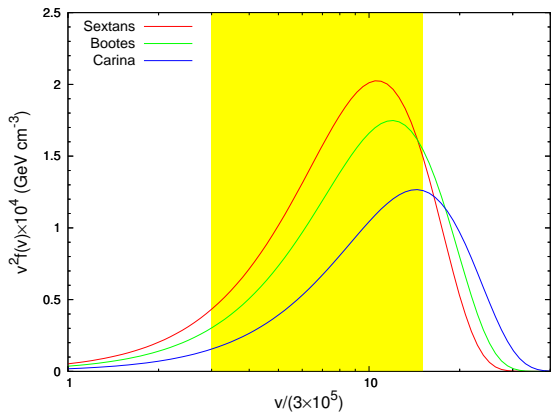}
\caption{Plots of $v^2f(v)$ as a function of velocity, evaluated at radius $r=0.5r_s$. The yellow band stands for the velocity of dark matter particle of $(3-15)\rm\;km/s$.}
\label{fig1}
\end{figure}
%*************************************************************************
%The velocity distribution function $f(r,v)$ depends on the NFW parameters $(r_s,\rho_{s})$. 
In figure 1, we plot $v^2f(r,v)$ at radius $r=0.5r_s$ as a function of dark matter particle velocity (in unit c). 
Results for the dwarf galaxy Sextans, Bootes and Carina are shown by the red, green and blue lines. We take the astrophysical parameters ($V_{\rm max},\;R_{V_{\rm max}}$) for these dwarf galaxies from Ref. \cite{Martinez2015}, and the NFW profile parameters $(r_s,\rho_{s})$ are determined using Eq. (2.4) and Eq. (2.5) (see the next subsection).
The measurements of the dwarf galaxy stellar velocity dispersion profiles show that the velocity typically spans 3 to 15 km/s (see figure 1 and figure 2 of Ref. \cite{Bonnivard2015}), this is represented by the yellow band in the figure.
%These results show that theoretical calculation of dark matter velocity distribution are in accordance with the measurements to a large extent.

%*************************************************************************
\subsection{Dark matter profile}

The spatial distribution of dark matter is assumed to be given by the NFW profile \cite{NFW1997}
\begin{eqnarray}
\rho(r)=\frac{\rho_s}{(r/r_{s})(1+r/r_{s})^2},
\end{eqnarray}
where $r_s$ and $\rho_s$ are the scale radius and density of the dark matter distribution profile.

For the subhalos, the scale parameters $r_s$ and $\rho_s$ are uniquely determined for each subhalo with the relations
\begin{eqnarray}
r_s&=&\frac{R_{V_{\rm max}}}{2.163}\\
\rho_s&=&\frac{4.625}{4\pi G}\left(\frac{V_{\rm max}}{r_s} \right)^2,
\end{eqnarray}
where $r_s$ is in kpc, $V_{\rm max}$ and $R_{V_{\rm max}}$ are the maximum circular velocity and the radius of maximum circular velocity.
Following Ref. \cite{Ackermann2012}, the relations between $V_{\rm max}$, $R_{V_{\rm max}}$ and the subhalo mass $M$ can be parameterized as
\begin{eqnarray}
V_{\rm max}&=&V_0\left ( \frac{M}{M_{\odot }} \right )^{\beta }\\
R_{V_{\rm max}}&=&R_0\left ( \frac{M}{M_{\odot }} \right )^{\delta },
\end{eqnarray}
where $V_0=10^{-1.20\pm 0.05}\rm\;km\;s^{-1}$, $\beta=0.30\pm 0.01$ and $R_0=10^{-3.1\pm 0.4}\rm\;kpc$ and $\delta=0.39\pm 0.02$, with a log-normal scatter $\sigma_{V_{\rm max}}=0.063\rm\;km\;s^{-1}$ and $\sigma_{R_{V_{\rm max}}}=0.136\rm\;kpc$ for each fitting equation.

We notice that although under the framework of $\Lambda$CDM cosmology, large N-body numerical simulations lead to the commonly used NFW halo cusp spatial density profile, 
analysis of observations in the central regions of various dwarf halos is in favor of cored profiles \cite{Kuhlen2012, Diemand2011, Frenk2012, Blok2010}. The self-interacting dark matter scenario \cite{Spergel2000} may provide one of the possible solutions to this problem (recent researches of the self-interacting dark matter with N-body simulations can be found in Refs. \cite{Vogelsberger2012, Zavala2013, Rocha2013, Peter2013}). In this scenario, the dark matter distribution may be altered intensely by the dark matter self-interaction and tends to a flat dark matter density profile in the center region of the halo (the reader may see Refs. \cite{Kaplinghat2016, Tulin2013, Kamada2017} for more details).

%*************************************************************************

\section{Sommerfeld enhancement}

The ``Sommerfeld enhancement" is the effect that enhances the dark matter annihilation cross sections in the low-velocity regions, this nonrelativistic quantum effect arises due to the formation of long range attractive interaction. If the dark matter particles interact via the exchange of some kind of light mediator, their incoming wave function can be distorted by the presence of a long range potential when their kinetic energy is low enough \cite{Hisano2005, Feng2010-2}. 
In view of the quantum field theory, the Sommerfeld effect corresponds to the contribution of ``ladder" Feynman diagrams like the one shown in figure 2, in which the force carrier is exchanged many times before the annihilation finally occurs \cite{Lattanzi2009}. This process gives rise to nonperturbative corrections to the annihilation cross section of dark matter\cite{Hisano2005}. 

We consider a hidden sector in which the dark matter particle $\chi$ couples to a light force carrier $\phi$ with coupling $g_{X}=\sqrt{4\pi\alpha_{X}}$. The actual annihilation cross section times relative velocity is $(\sigma_{\rm ann}v_{\rm rel})=(\sigma_{\rm ann}v_{\rm rel})_{0}\times S$, where $(\sigma_{\rm ann}v_{\rm rel})_{0}$ is the tree-level cross section times relative velocity and $S$ stands for the Sommerfeld enhancement factor. We assume that the dark matter tree-level annihilation cross section is dominated by $s$-wave processes, which are unsuppressed at low velocities .
Let $\psi(r)$ be the reduced two-body wave function of the $s$-wave annihilation, in the nonrelativistic limit, the motion of the reduced two-body wavefunction obeys the radial Schr\"odinger equation
\begin{eqnarray}
\frac{1}{m_{\chi}}\frac{d^2\psi}{dr^2}-V(r)\psi(r)=-m_{\chi}v^{2}\psi(r),
\end{eqnarray}
where $m_{\chi}$ is the dark matter mass and $v=v_{\rm rel}/2$ is the velocity of each particle in the center-of-mass frame. The attractive force of nonrelativistic dark matter particles mediated by a scalar or vector forms a Yukawa potential
\begin{eqnarray}
V(r)=-\frac{\alpha_{X}}{r}e^{-m_{\phi}r},
\end{eqnarray}
where $m_{\phi}$ is the mass of the light mediator. By solving the Schr\"odinger equation with the boundary condition ${\psi }'(r)=im_{\chi}v\psi(r)$ and $\psi(r)=e^{im_{\chi}vr}$ as $r\to \infty$, we obtain the Sommerfeld enhancement factor
\begin{eqnarray}
S=\left | \frac{\psi(\infty)}{\psi(0)} \right |^2.
\end{eqnarray}
Defining two dimensionless parameters
\begin{eqnarray}
\varepsilon_{v}=\frac{v}{\alpha_{X}} \quad {\rm and} \quad \varepsilon_{\phi}=\frac{m_{\phi}}{\alpha_{X}m_{\chi}},
\end{eqnarray}
and using the dimensionless variable $x=\alpha_{X}m_{\chi}r$, the radial Schr\"odinger equation can be rewritten as 
\begin{eqnarray}
{\psi }''(x)+[\varepsilon_{v}^2+V(x)]\psi(x)=0,
\end{eqnarray}
where the potential $V(x)=\exp(-\varepsilon_{\phi}x)/x$. The boundary conditions become ${\psi }'(x)=i\varepsilon_{v}\psi(x)$ and $\psi(x)=\exp(i\varepsilon_{v}x)$ as $x\to \infty$.
Approximating the Yukawa potential as the Hulth$\acute{\rm e}$n potential we can obtain an analytic formula for Sommerfeld enhancement \cite{Cassel2010, Slatyer2010}
\begin{eqnarray}
S=\frac{\pi}{\varepsilon_{v}}\frac{\sinh\left ( \frac{2\pi\varepsilon_{v}}{\pi^{2}\varepsilon_{\phi}/6} \right )}{\cosh\left ( \frac{2\pi\varepsilon_{v}}{\pi^{2}\varepsilon_{\phi}/6} \right )-\cos\left ( 2\pi\sqrt{\frac{1}{\pi^{2}\varepsilon_{\phi}/6}-\frac{\varepsilon_{v}^{2}}{(\pi^{2}\varepsilon_{\phi}/6)^{2}}} \right )}.
\end{eqnarray}
In this work, we have used this analytic approximation to the Sommerfeld enhancement.
Comparisons between the analytic and numerical solutions for the Sommerfeld enhancement are shown in appendix B. 
 
%*****************************fig2***************************************
\begin{figure}
\centering
\includegraphics[width=70mm,angle=0]{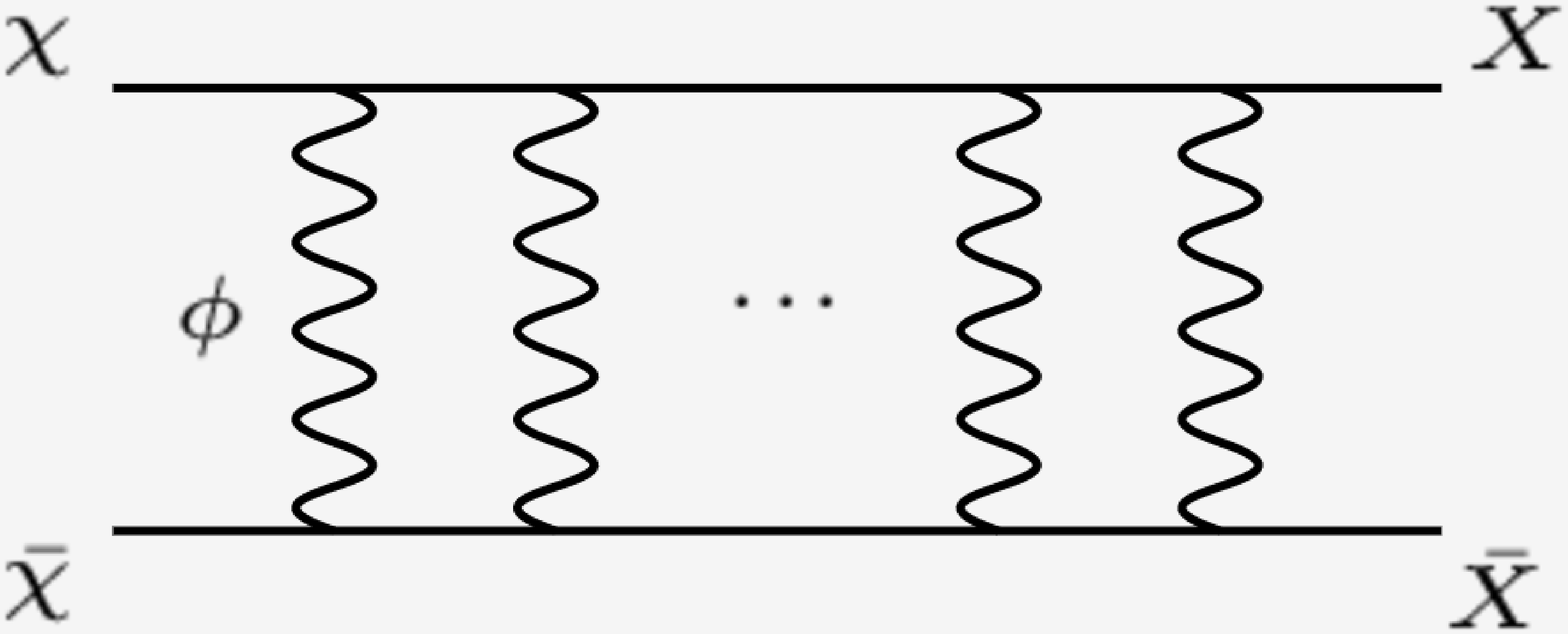}
\label{fig2}
\caption{Feynman diagram giving rise to the Sommerfeld enhancement for $\chi\bar{\chi}\to X\bar{X}$ annihilation processes, wave line stands for the exchange of a light mediator.}
\end{figure}

%*************************************************************************

%********************************************************************************************************

\section{Sommerfeld-enhanced $J$-factor}

The commonly used differential photon flux produced from the annihilation of astrophysical dark matter is given by
\begin{eqnarray}
\frac{d\Phi }{dE_{\gamma}}&=&\frac{1}{4\pi}\frac{\left (\sigma_{\rm ann}v_{\rm rel} \right )_0}{2m_{\chi}^2}\frac{dN}{dE_{\gamma}}\times \int_{\Delta \Omega }d\Omega\int_{\rm LOS}dl\rho^{2}_{\chi}[r(D,l,\theta)]\\
&=&\frac{1}{4\pi}\frac{\left (\sigma_{\rm ann}v_{\rm rel} \right )_0}{2m_{\chi}^2}\frac{dN}{dE_{\gamma}}\times J,
\end{eqnarray}
where $D$ is the distance to the satellite center, $l$ is the distance along the line of sight, $\theta$ is the offset angle relative to the center and $r(D,l,\theta)=\sqrt{D^2+l^2-2Dl\cos\theta}$ is the distance from the satellite center. The angular integration is performed over $\Delta \Omega =2\pi(1-\cos\theta)$. $dN/dE_{\gamma}$ is the photon spectrum produced by a single annihilation process and $\left (\sigma_{\rm ann}v_{\rm rel} \right )_0$ is the dark matter annihilation cross section. The second term of Eq. (4.1) is often called $J$-factor, which encapsulates all astrophysical information for determining the photon flux from the dark matter annihilation.

Notice that the annihilation cross section is assumed to be velocity-independent to obtain the differential photon flux Eq. (4.1), thus the $J$-factor is simply an integral over the line-of-sight and over a given angular region of the square of the dark matter density profile of the satellite. This is reasonable for the most discussed neutralino models featuring a mostly used $s$-wave velocity-independent cross section. However, theoretically, the annihilation cross section may be velocity-dependent. For instance, dark matter models with $p$-wave suppressed annihilation processes have been studied in many recent works \cite{Diamanti2014, Berlin2014-1, lu2016-1}. Additionally, dark matter models with Sommerfeld-enhanced annihilation cross section have been used to account for the observed positron cosmic-ray anomalies \cite{Hamed2009, Feng2010-1, Feng2010-2, Bergstrom2009, Ibarra2014}.
In these cases, the $J$-factor cannot be simply factorized from the particle physics, since the photon flux arising from dark matter annihilation depends on the dark matter velocity distribution.
A general form of differential photon flux produced from dark matter with a velocity-dependent annihilation cross section can be written as
\begin{eqnarray}
\frac{d\Phi }{dE_{\gamma}}=\frac{1}{4\pi}\frac{1}{2m_{\chi}^2}\frac{dN}{dE_{\gamma}}\times \int_{\Delta \Omega }d\Omega\int_{\rm LOS}dl\int d^3\vec{v}_{1}f(r,\vec{v}_{1})\int d^3\vec{v}_{2}f(r,\vec{v}_{2})\left (\sigma_{\rm ann}v_{\rm rel} \right ).
\end{eqnarray}
In this formula, the annihilation cross section $\left (\sigma_{\rm ann}v_{\rm rel} \right )$ may be $p$-wave suppressed or Sommerf eld-enhanced. In this work we only take into account the Sommerfeld enhancement which is dominated by $s$-wave processes, the calculations of Sommerfeld-enhanced $J$-factor are based on the methods given by Boddy et al. \cite{Boddy2017}.
Notice that the Sommerfeld-enhanced annihilation cross section can be expressed as $(\sigma_{\rm ann}v_{\rm rel})=(\sigma_{\rm ann}v_{\rm rel})_{0}\times S(v_{\rm rel}/2)$ \cite{Cirelli2007, Hamed2009}, where the tree-level cross section $(\sigma_{\rm ann}v_{\rm rel})_{0}$ is $s$-wave dominant, i.e., velocity-independent. 
%The condition $(\sigma_{\rm ann}v_{\rm rel})_{0}=3\times 10^{-26}\rm\;cm^{2}s^{-1}$ is adopted here to reproduce the observed relic abundance.
Then the Eq. (4.3) can be rewritten as
\begin{eqnarray}
\frac{d\Phi }{dE_{\gamma}}=\frac{1}{4\pi}\frac{\left (\sigma_{\rm ann}v_{\rm rel} \right )_{0}}{2m_{\chi}^2}\frac{dN}{dE_{\gamma}}\times J_{S},
\end{eqnarray}
where $J_{S}$ is the Sommerfeld-enhanced $J$-factor which is characterized by a subscript $S$
\begin{eqnarray}
J_{S}&=&\int_{\Delta \Omega }d\Omega\int_{\rm LOS}dl\int d^3\vec{v}_{1}f(r,\vec{v}_{1})\int d^3\vec{v}_{2}f(r,\vec{v}_{2})S(\left |\vec{v}_{2}-\vec{v}_{1}\right |/2)\\
&\simeq& \frac{1}{D^2}\int dV\int d^3\vec{v}_{1}f(r,\vec{v}_{1})\int d^3\vec{v}_{2}f(r,\vec{v}_{2})S(\left |\vec{v}_{2}-\vec{v}_{1}\right |/2).
\end{eqnarray}
In the limit $S\to 1$ (i.e., $\varepsilon_{\phi }\gg 1$), we have $J_S\to J$. The Sommerfeld-enhanced $J$-factor $J_{S}$ depends on five parameters, two of them are dark matter model parameters: $\alpha_{X}$ and $\varepsilon_{\phi}$, and the other three are astrophysical parameters: $(r_s,\rho_{s})$, the scales of satellite dark matter profile and $D$, the distance to the satellite center.

Notice that the integration of $J_{S}$ not only depends on the magnitudes of $v_{1}$ and $v_{2}$ but also depends on the angle between the velocity of two dark matter particles, we express this explicitly
\begin{eqnarray}
\begin{aligned}
J_{S}=\frac{32\pi^3}{D^2}\int_{0}^{r_{\rm max}}r^2dr\int_{0}^{v_{\rm max}}v_1^2f(r,v_1)dv_1\int_{0}^{v_{\rm max}}v_2^2f(r,v_2)dv_2\\
\times \int_{0}^{\pi}\sin\theta S\left (\sqrt{v_1^2+v_2^2-2v_1v_2\cos\theta}/2\right )d\theta.
\end{aligned}
\end{eqnarray}

%*****************************fig3***************************************
\begin{figure}
\centering
\includegraphics[width=60mm,angle=0]{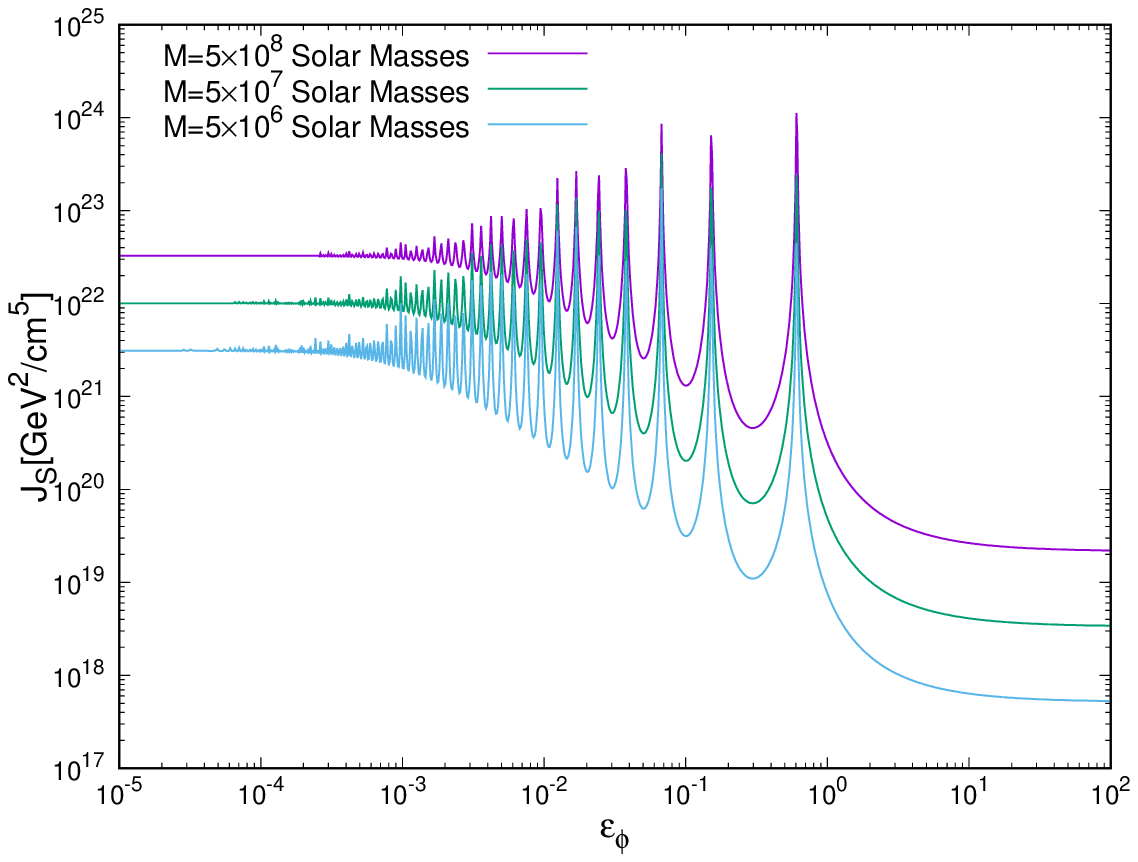}
\includegraphics[width=60mm,angle=0]{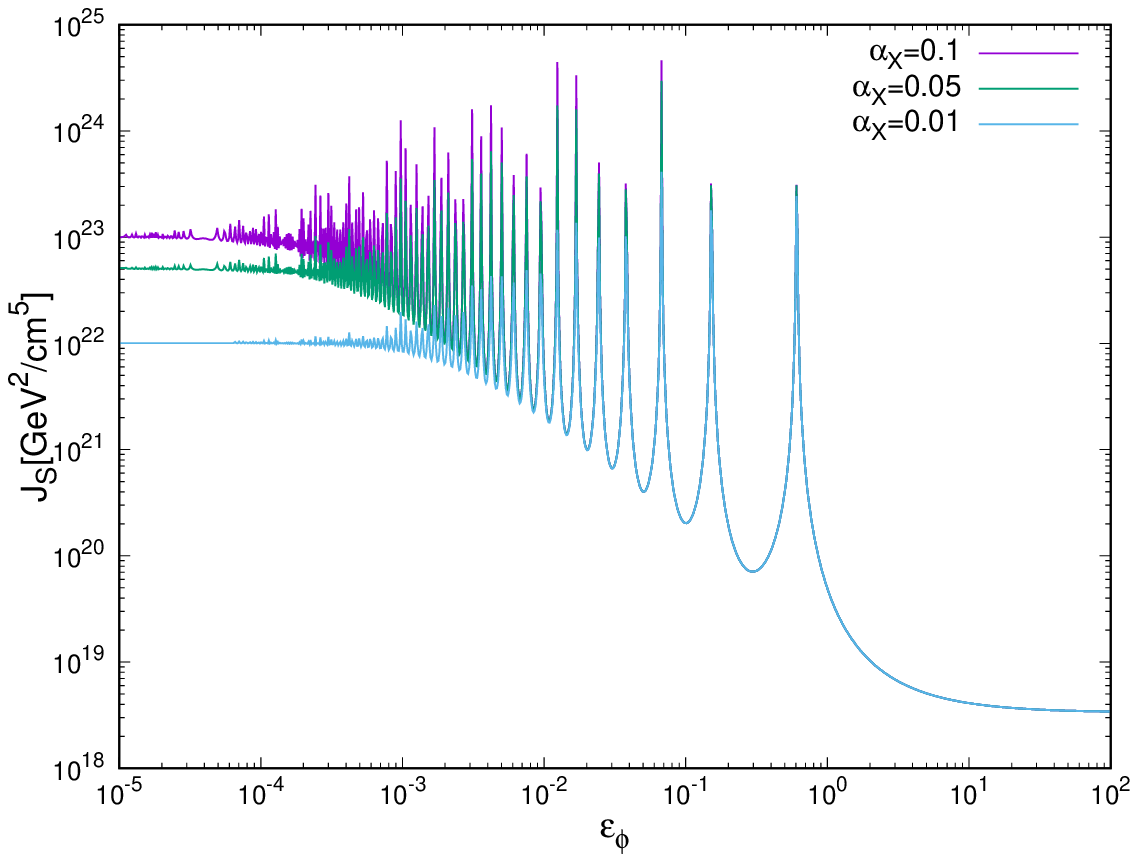}
\caption{Left panel: Sommerfeld-enhanced $J$-factor as a function of $\varepsilon_{\phi}$, assuming $D=50\rm\;kpc$ and $\alpha_{X}=0.01$. The purple, green and blue line stand for the results with the subhalo masses $M=5\times 10^8\;M_{\odot }$, $5\times 10^7\;M_{\odot }$ and $5\times 10^6\;M_{\odot }$. Right panel: Sommerfeld-enhanced $J$-factor as a function of $\varepsilon_{\phi}$, assuming $D=50\rm\;kpc$ and $M=5\times 10^7\;M_{\odot }$. The purple, green and blue line represent the results with the $\alpha_{X}=0.1$, $0.05$ and $0.01$.}
\label{fig3}
\end{figure}
%*************************************************************************

$J_{S}$ is depicted in figure 3 as a function of $\varepsilon_{\phi}$ with various parameters. The distance $D$ in the two panels of the figure is assumed to be $50\rm\;kpc$. In the left panel of figure 3, we calculate the Sommerfeld-enhanced $J$-factor for three mass scales of subhalo: $5\times 10^6\;M_{\odot }$, $5\times 10^7\;M_{\odot }$ and $5\times 10^8\;M_{\odot }$, with the coupling $\alpha_{X}=0.01$. We obtain the scales of dark matter profile $(r_s,\rho_{s})$ for the given subhalo masses using the relations mentioned in the above section, and notice that there are some uncertainties in the fitting equations.
In the right panel, we demonstrate the Sommerfeld-enhanced $J$-factor with various coupling $\alpha_{X}$, assuming $M=5\times 10^7\;M_{\odot }$ and $D=50\rm\;kpc$. As is shown in the figure, the Sommerfeld-enhanced $J$-factor $J_{S}\to J$ as $\varepsilon_{\phi}$ goes much beyond 1, since in this region the Sommerfeld enhancement $S\to 1$. The enhanced factor $J_S/J$ may be up to $\sim 10^3$ for $\alpha_{X}=0.01$ at low $\varepsilon_{\phi}$.
The resonances occur at $\varepsilon_{\phi}\simeq \frac{6}{\pi^2n^2}$, where $n$ is an integer, and the Sommerfeld enhancement scales as $\alpha_{X}^2\varepsilon _{\phi}/v^2$. The resonance amplitude decreases with the increasing of a subhalo mass, since the average speed of dark matter particles is larger in a much more massive subhalo. The saturation of the Sommerfeld enhancement occurs at about 
$\varepsilon_{\phi}\sim \frac{v}{2\alpha_{X}}$, this leads to the constraint $\varepsilon_{\phi}\gtrsim 10^{-5}\alpha_{X}^{-1}$ for the dwarf galaxies, in which the typical dark matter velocity has the scale $\sim\mathcal{O}(10)\rm\;km/s$.
As is mentioned above, the Sommerfeld enhancement $S\to \pi\alpha_{X}/v$ in the limit $\varepsilon_{\phi}\ll \varepsilon_v$ (i.e., $\varepsilon_{\phi}\lesssim 10^{-4}\alpha_{X}^{-1}$ for the dwarf galaxies). Thus in this region the Sommerfeld-enhanced $J$-factor is proportional to the coupling $\alpha_{X}$, as is shown in the figure. For the region $\varepsilon_{\phi}\gtrsim 10^{-4}\alpha_{X}^{-1}$, the Sommerfeld-enhanced $J$-factor is independent of $\alpha_{X}$.

In order to focus on our topic, we have used the NFW profile and assumed isotropic orbits for the calculation of $J$-factor. Notice that the previous literatures \cite{Bonnivard2015, Martinez2015, Sameth2015} about the $J$-factor calculations have allowed for non-NFW profiles, anisotropic stellar velocity dispersions and assumed a Gaussian likelihood for the stellar velocities. Refs. \cite{Sanders2016, Evans2016} provide a simple analytic formula to calculate the $J$-factor for spherical cusp without solving the spherical Jeans equations, which relate the velocities of the stars to the underlying dark matter distribution. In the framework of isotropic orbits, the calculations of non-NFW profiles are analogous, however, it may require to go beyond the Eddington's formula when the anisotropic models are taken into account.

\section{Constraints from the Fermi-LAT gamma-ray sources observations}

\subsection{Constraints from the Fermi-LAT subhalo searches}

Since 2010, the Fermi collaboration has released their First Fermi-LAT catalog (1FGL) \cite{1FGL}, based on their first 11 months of data. This catalog contains 1451 sources detected and characterized in the 100 MeV to 100 GeV range, 630 of which had (at the time) not been associated with counterparts at other wavelengths.
Recently, Bertoni et al. \cite{Berlin2014, Bertoni2015, Hooper2017} examined the subset of 992 unassociated sources in the Third Fermi-LAT catalog (3FGL) \cite{3FGL} in an effort to identify dark matter subhalo candidates, and to use the population of such sources to constrain the dark matter annihilation cross section. They identified a subset of 24 bright ($\Phi_{\gamma}>7\times 10^{-10}\rm\;cm^{-2}\;s^{-1}$, $E>1\rm\; GeV$) and high-latitude ($|b| > 20^{\circ}$) sources (a list of these sources can be found in table 1 of Ref. \cite{Bertoni2015}) that show no evidence of variability and exhibit a spectral shape that is consistent with the predictions of annihilating dark matter.

The differential gamma-ray spectrum from dark matter annihilation within an individual subhalo is given by Eq. (4.1) for a velocity-independent cross section and Eq. (4.3) for a velocity-dependent cross section. For the case of velocity-independent dark matter annihilation cross section, the brightness of the gamma rays from a subhalo mainly depends on the mass and distance of the subhalo and also the dark matter density profile.
Throughout this work, we use the NFW profile for the dark matter distribution of subhalos, however, we notice that for subhalos located within the innermost few tens of kiloparsec of their host halo, the distribution of subhalos is significantly altered by the tidal effects, which lead to a power-law profile with an exponential cutoff (see Ref. \cite{Hooper2017} for more details).

%*****************************fig4***************************************
\begin{figure}
\centering
\includegraphics[width=60mm,angle=0]{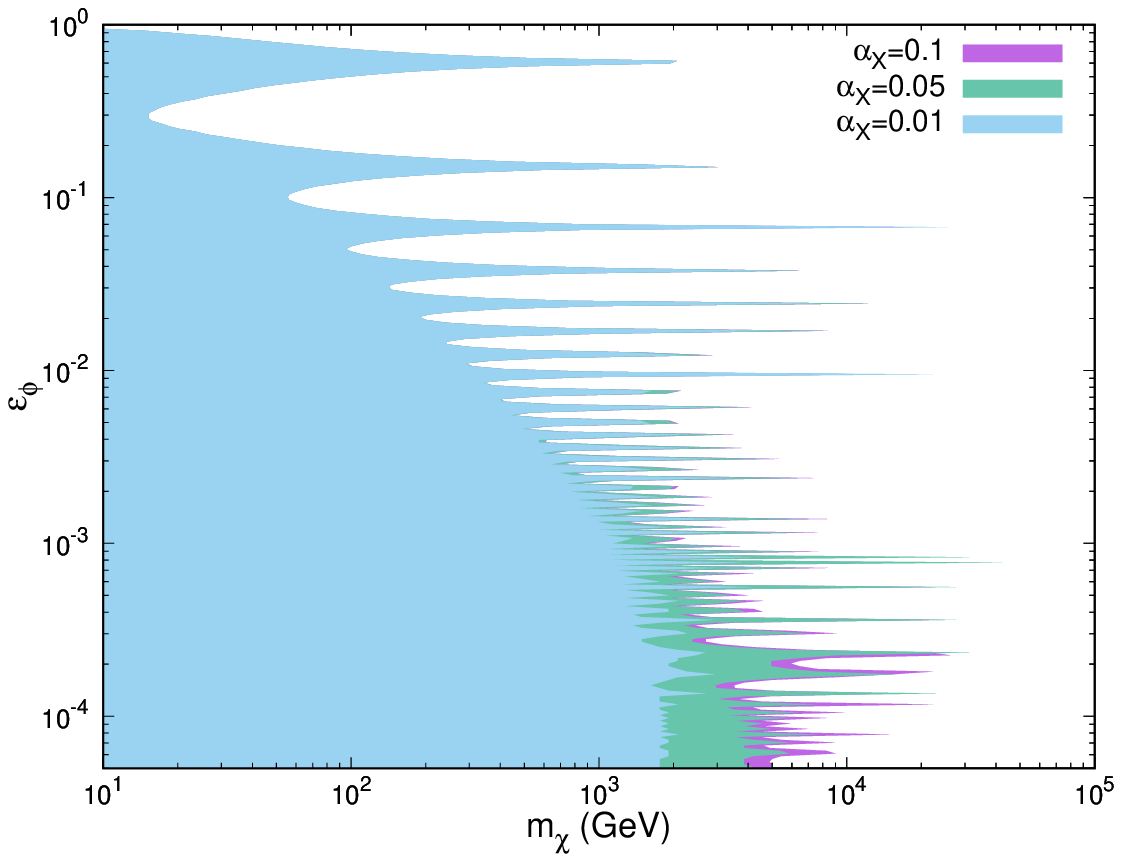}
\includegraphics[width=60mm,angle=0]{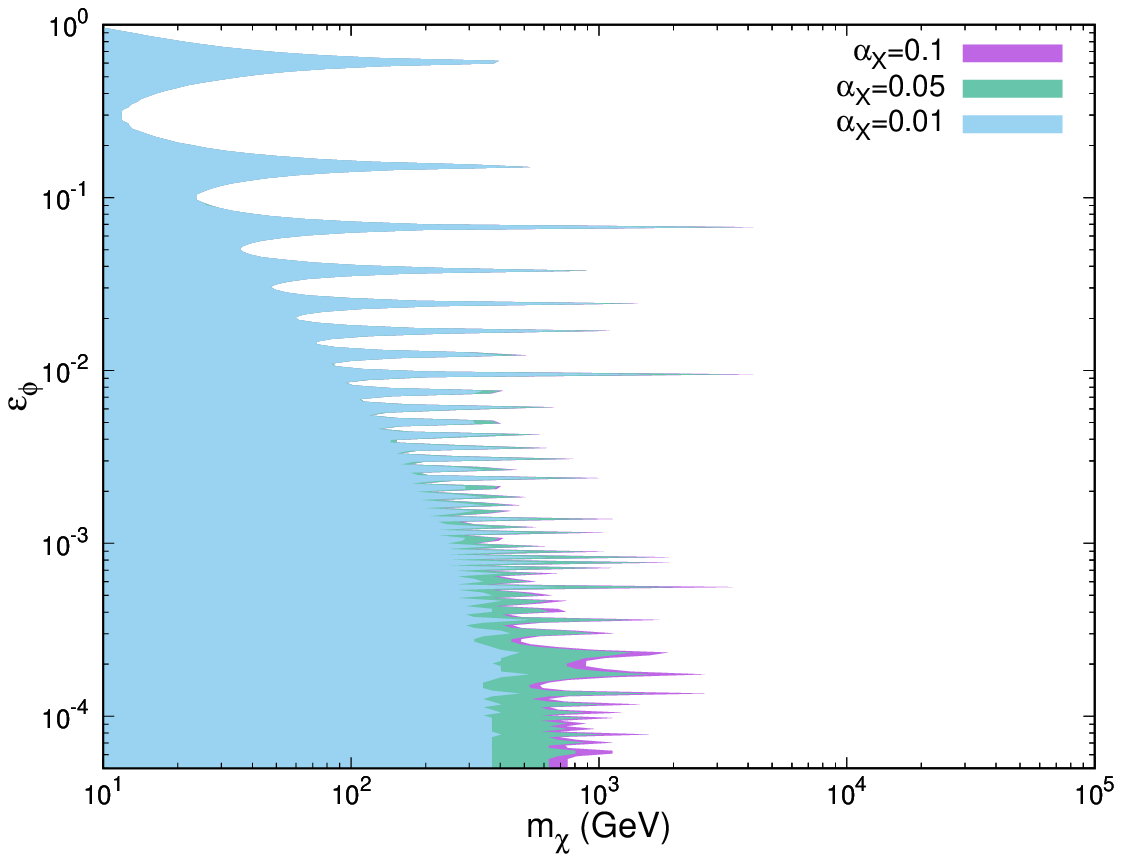}
\caption{The Fermi-LAT subhalo observation results constrain on the $\varepsilon_{\phi}$ at 95\% confidence level. The color shaded parameter space is excluded at 95\% confidence level, the dark matter
annihilation channel is assumed to be $b\bar{b}$ and $\tau\bar{\tau}$ in the left and right panel.}
\label{fig4}
\end{figure}

%*************************************************************************
%*****************************fig5***************************************
\begin{figure}
\centering
\includegraphics[width=60mm,angle=0]{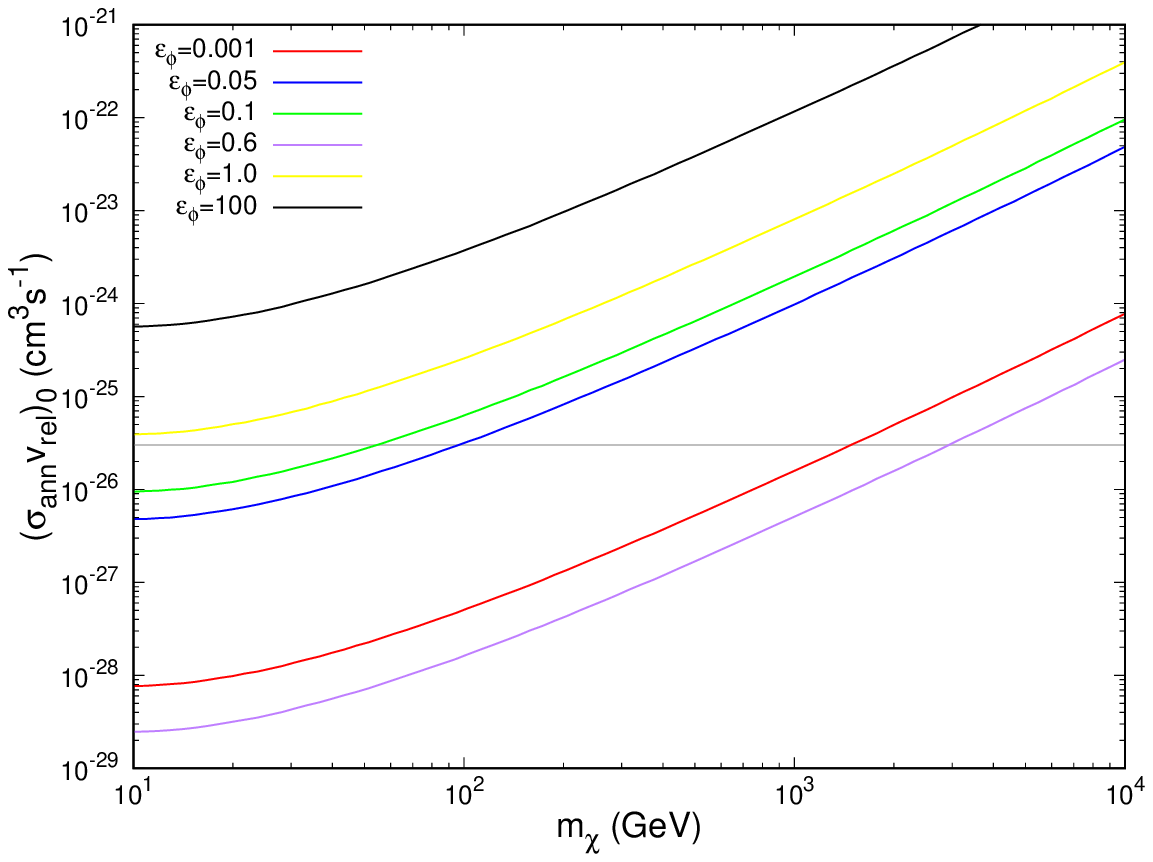}
\includegraphics[width=60mm,angle=0]{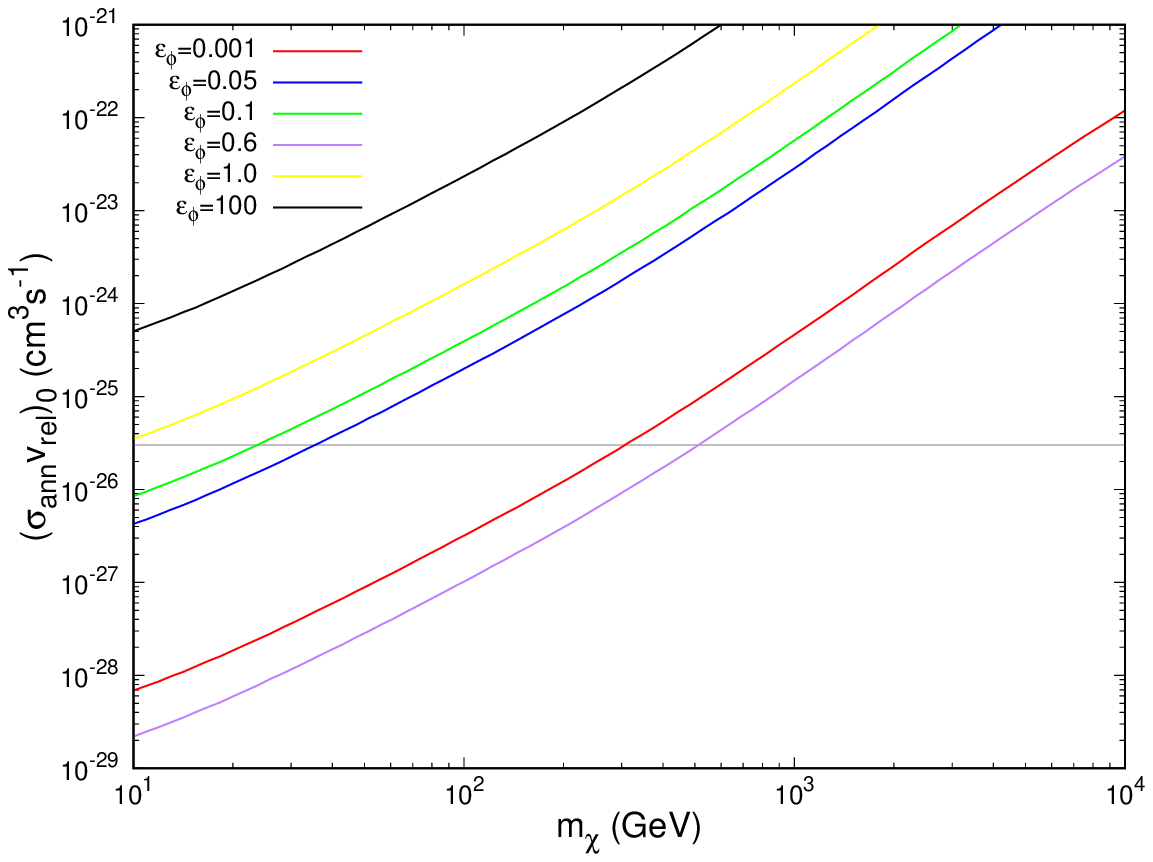}
\caption{Constraints on the dark matter annihilation cross section at 95\% confidence level derived from the Fermi-LAT subhalo searches, with a given Sommerfeld-enhanced $J$-factor. The horizontal grey line stands for an annihilation cross section $(\sigma_{\rm ann}v_{\rm rel})_0=3\times 10^{-26}\rm\;cm^{3}\;s^{-1}$, the dark matter annihilation channel is assumed to be $b\bar{b}$ and $\tau\bar{\tau}$ in the left and right panel.}
\label{fig5}
\end{figure}

%*************************************************************************
%*****************************fig6***************************************
%\begin{figure}
%\centering
%\includegraphics[width=60mm,angle=0]{./ml_100.eps}
%\includegraphics[width=60mm,angle=0]{./ml_005.eps}
%\caption{The 95\% confidence level upper limits on the dark matter annihilation cross section. The red, green and blue lines stand for the constraints with minimum subhalo masses $1M_{\odot}$, $10^{3}M_{\odot}$ and $10^{5}M_{\odot}$, $\varepsilon_{\phi}$ is set at 100 and 0.05 in the left and right panel.}
%\label{fig6}
%\end{figure}

%*************************************************************************

Cosmological simulations \cite{VLII2008, ELVIS2013} show that the mass distribution of subhalo populations follows a power-law form, $dN/dM\propto M^{-1.9}$. We make use of the mass distribution of the local subhalo population from Ref. \cite{Hooper2017}, in which the distribution is obtained through fitting to the ELVIS simulation data with subhalo masses between $10^{8}M_{\odot}$ and $10^{10}M_{\odot}$
\begin{eqnarray}
\frac{dN}{dMdV}=\frac{628}{M_{\odot }\rm kpc^{3}}\left ( \frac{M}{M_{\odot}} \right )^{-1.9}.
\end{eqnarray}
For subhalo masses lower than $\sim 10^{8}M_{\odot}$, the numbers of subhalos of the simulation are smaller than that predicted by the power-law formula, this may be due to the finite resolution of the simulation \cite{Hooper2017}. Larger numbers of subhalos are expected when taking into account the finite resolution of the simulation and the power-law form may still be valid for the subhalo mass distribution in these mass ranges.
Thus, following Ref. \cite{Hooper2017}, in this work we use the power-law form of subhalo mass distribution for the masses of subhalo in the range of $(1.9\times 10^{5}-10^{7})M_{\odot}$.
The lower limit $1.9\times 10^{5}M_{\odot}$ of subhalo masses represents the resolution of the ELVIS simulation, and the upper limit of subhalo masses is restricted to be below $10^{7}M_{\odot}$ to avoid the inclusion of any dwarf galaxies.
There are no observation constraints on the minimum masses of subhalo, the limits on the dark matter annihilation cross section may alter by a factor of $\lesssim 2$ with the variation of the minimum subhalo masses.

With the subhalo mass distribution, the predicted number of dark matter subhalos that yield a gamma-ray flux above the Fermi-LAT threshold $\Phi_{\rm threshold}$ is given by
\begin{eqnarray}
N_{\rm pre}=\Omega \int\int \frac{dN}{dMdV}D^{2}\Theta(\Phi_{\gamma}-\Phi_{\rm Thresh})dMdD,
\end{eqnarray}
where $\Theta$ is the step function, $\Phi_{\gamma}$ is the gamma-ray flux produced from a subhalo with energies in $(1-100)\rm\;GeV$, and $\Omega$ is the solid angle observed by the Fermi experiment, which corresponds to $4\pi(1-\sin 20^{\circ})$ for the case of $|b|>20^{\circ}$. We take the threshold $\Phi_{\rm threshold}$ as that in Ref. \cite{Hooper2017}.

Using the method proposed in previous works \cite{Berlin2014, Bertoni2015, Hooper2017}, we put constraints on the Sommerfeld enhancement with the subhalo observations. 
Given a subhalo mass, we calculate the maximum circular velocity $V_{\rm max}$ and the radius of maximum circular velocity $R_{V_{\rm max}}$ with Eq. (2.6) and Eq. (2.7). The subhalo NFW profile parameters $r_{s}$ and $\rho_{s}$ are determined using Eq. (2.4) and Eq. (2.5). We then determine the dark matter particle velocity distribution function $f(r,v)$ using Eq. (A.5) and Eq. (A.7) (more details on the calculations of $f(r,v)$ can be found in appendix A). 
Given the dark matter particle parameters $\alpha_{X}$, $\varepsilon_{\phi}$ and annihilation cross section, the Sommerfeld-enhanced $J$-factor of subhalo at distance $D$ and the differential photon flux produced from dark matter annihilation are determined using Eq. (4.7) and Eq. (4.4). By integrating the differential photon flux $d\Phi_{\gamma}/dE_{\gamma}$ with respect to photon energy $E_{\gamma}$ we obtain the gamma-ray flux $\Phi_{\gamma}$ (following Refs. \cite{Berlin2014, Bertoni2015, Hooper2017}, we take into account photons with energy in the range of (1-100) GeV).
With these results, we finally obtain the predicted number of dark matter subhalos $N_{\rm pre}$ that may be observed by the Fermi-LAT experiment using Eq. (5.2).
For the case of small numbers of observed events, the confidence limits can be placed based on the Poisson statistics. 
Based on the results from the analysis by Bertoni et al. in Ref. \cite{Bertoni2015}, the number of dark matter subhalo candidates is taken to be 24, thus the 95\% confidence level (CL) Poisson upper limit on the predicted number of such sources is $N_{\rm upp}=33.75$ (see table 1 in Ref. \cite{Gehrels1986}).

In figure 4, we show the limits on the parameter $\varepsilon_{\phi}$ as a function of dark matter mass $m_{\chi}$ for several values of the coupling $\alpha_{X}$, and the dark matter cross section $(\sigma_{\rm ann}v_{\rm rel})_0$ is fixed at $3\times 10^{-26}\rm\;cm^{3}\;s^{-1}$.
For each choice of $\varepsilon_{\phi}$, we calculate the predicted number of sources $N_{\rm pre}$ that may be observed by the Fermi-LAT experiment. By plotting the contour of predicted number of sources at $N_{\rm pre}=N_{\rm upp}=33.75$ in the $(m_{\chi}-\varepsilon_{\phi})$ plane, we determine the 95\% confidence level limits on the parameter $\varepsilon_{\phi}$.
%, which corresponds to the Poisson upper limits on the number of sources. 
In the left and right panel of figure 4, the dark matter annihilation channel is $b\bar{b}$ and $\tau\bar{\tau}$ respectively.
The color shaded parameter spaces are excluded at 95\% confidence level by the subhalo observations. As is shown in the figure, the constraints become more stringent with a larger $\alpha_{X}$ for $\varepsilon_{\phi}\gtrsim 10^{-3}$. Our results can be easily changed to the $(m_{\phi}-m_{\chi})$ plane by using the relation $\varepsilon_{\phi}=m_{\phi}/(\alpha_{X}m_{\chi})$.

In figure 5, we show the constraints on the dark matter annihilation cross section at 95\% confidence level as a function of dark matter mass. We assume the coupling $\alpha_{X}$ to be $0.1$ and determine the Sommerfeld-enhanced $J$-factor for a given value of $\varepsilon_{\phi}$. We calculate the signal gamma-ray flux from dark matter annihilation for each choice of dark matter mass $m_{\chi}$ and annihilation cross section $(\sigma_{\rm ann}v_{\rm rel})_0$. Using the same procedure as above, we count the predicted numbers of sources and determine the 95\% confidence level Poisson upper limits on the dark matter annihilation cross section by plotting the contour of predicted number of sources at $N_{\rm pre}=N_{\rm upp}=33.75$.
Although the constraints derived here are somewhat weaker than those based on the observations of dwarf galaxies (see the next subsection), the results may be improved with more observations from the Fermi-LAT experiments and future gamma-ray telescopes.

Here we give discussions on some of the important uncertainties involved in our calculations and estimate their likely impacts on our limits. One of the uncertainties comes from the profile forms that might describe the distributions of dark matter in the subhalos. We have assumed the NFW dark matter profile for this study, the constraints may be enhanced by a factor of a few for the Einasto profile. We have taken a power-law form with a slope 1.9 for the dark matter mass distribution, as is used in Ref. \cite{Berlin2014, Bertoni2015, Hooper2017}. The limits on the dark matter cross section may alter by a factor of 1.5-2 when we change the slope over the range of 1.8 to 2.0 (see Ref. \cite{Hooper2017}).
We have assumed a subhalo mass range that begins at $\sim 10^{5}M_{\odot}$, which is the resolution of the ELVIS simulation. We have checked that the constraints remain unchanged if the minimum subhalo mass is set at the resolution of the Via Lactea II simulation $\sim 10^{3}M_{\odot}$. However, the limits may improve by a factor of $\sim 2$ when the minimum mass is taken to be $1M_{\odot}$.
%In figure 6, we assess the uncertainty associated with the minimum subhalo masses. Here we have adopted a value of 100 and 0.05 for $\varepsilon_{\phi}$ in the left and right panel of the figure and considered the case of dark matter annihilations to $b\bar{b}$. In each panel, we plot the upper limits on the dark matter cross section with three different values of minimum subhalo masses $M_{\rm min}=1\;M_{\odot}$ (red line), $M_{\rm min}=10^{3}\;M_{\odot}$ (green line), and $M_{\rm min}=10^{5}\;M_{\odot}$ (blue line). As is shown in the figure, the limits do not alter when the minimum mass increases from $10^{3}\;M_{\odot}$ to $10^{5}\;M_{\odot}$, however, the limits are improved by a factor of $\sim 2$ when the minimum mass is taken to be $1\;M_{\odot}$. The results also show that the uncertainty from minimum subhalo masses does not depend on the value of $\varepsilon_{\phi}$.

\subsection{Constraints from the Fermi-LAT dSphs searches}

One of the most stringent constraints on the dark matter annihilation cross section to date comes from gamma-ray observations of Milky Way dwarf spheroidal satellite galaxies (dSphs).
Here we take advantage of the Fermi-LAT dark matter searches from Milky Way dSphs to put limits on the Sommerfeld enhancement.
For this purpose, we follow the results from the analysis by the Fermi collaboration \cite{Fermi2015}, which are based on the six years of Fermi-LAT gamma-ray data observation on the 15 dwarf galaxies in the energy range $500\rm\;MeV-500\;GeV$. 
%We perform the statistical analysis using the likelihood in each energy bin as a function of integrated signal flux provided by the Fermi collaboration \cite{Fermi2015}.

Firstly, we will show how the $J$-factor uncertainty impacts the constraints of the dark matter annihilation cross section. 
Bin-by-bin likelihood $\mathcal{L}_{i}(\bm{\mu},\bm{\theta}_{i}|\mathcal{D}_{i})$ as a function of gamma-ray flux for 15 dSphs is available online \cite{online} provided by the Fermi collaboration \cite{Fermi2015}.
The total likelihood function for target dwarf galaxy $i$ is given by
\begin{eqnarray}
\tilde{\mathcal{L}}_{i}(\bm{\mu},\bm{\theta}_{i}=\left \{ \bm{\alpha},J_{i} \right \}|\mathcal{D}_{i})=\mathcal{L}_{i}(\bm{\mu},\bm{\theta}_{i}|\mathcal{D}_{i})\mathcal{L}_{J}(J_{i}|J_{{\rm obs},i},\sigma_{i}),
\end{eqnarray}
where $\mathcal{L}_{i}(\bm{\mu},\bm{\theta}_{i}|\mathcal{D}_{i})$ and $\mathcal{L}_{J}(J_{i}|J_{{\rm obs},i},\sigma_{i})$ are the likelihood functions from the dark matter models and $J$-factors, $\bm{\mu}$ is the set of parameters of dark matter model, and $\bm{\theta}_{i}$ is the set of nuisance parameters (see Refs. \cite{Fermi2014, Fermi2015} for more details). 
Since the constraints on the dark matter annihilation cross section given by the Fermi collaboration \cite{Fermi2015} have already taken into account the $J$-factor uncertainty, we will calculate the constraints without consideration of this uncertainty. To do this, we set the $J$-factor $J_{i}=J_{{\rm obs},i}$ for each target (here the $J_{{\rm obs},i}$ is from Ref. \cite{Fermi2015}), and take $\mathcal{L}_{i}(\bm{\mu},\bm{\theta}_{i}|\mathcal{D}_{i})$ as the total likelihood function. Given the dark matter mass and annihilation cross section (here the annihilation cross section is velocity-independent), we compute the gamma-ray flux produced from dark matter annihilations in each energy bin and determine the likelihood $\mathcal{L}_{i}(\bm{\mu},\bm{\theta}_{i}|\mathcal{D}_{i})$ using the likelihood function tables provided by the Fermi collaboration online \cite{online}. Thus we obtain the likelihood in all energy bins for each target, multiply all of the likelihood functions we finally obtain the total likelihood from the 15 dSphs (i.e., $\mathcal{L}=\prod_{i}\mathcal{L}_{i}(\bm{\mu},\bm{\theta}_{i}|\mathcal{D}_{i})$).
We then determine the regions of the annihilation cross section excluded at 95\% confidence level by performing a test statistic, comparing the likelihood with and without the dark matter signal.
In figure 6, we show our calculation results (solid lines) and compare them with the constraints given by the Fermi collaboration (dot line). As we can see, the constraints can be enhanced at most $2\%-5\%$ when the $J$-factor uncertainty is not taken into account, and the enhancements go to zero when dark matter mass $m_{\chi} \gtrsim 1\rm\;TeV$. 
%*****************************fig6***************************************
\begin{figure}
\centering
\includegraphics[width=80mm,angle=0]{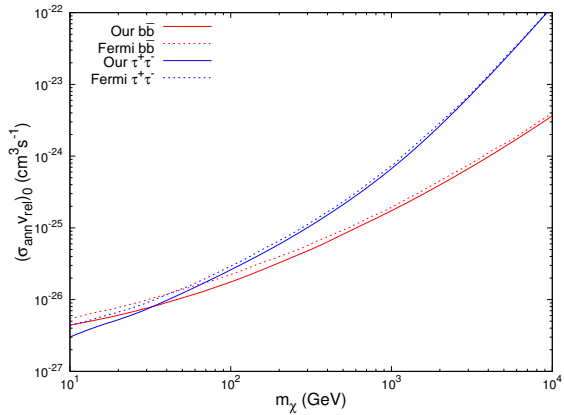}
\caption{Comparison of our calculation results without taking into account the $J$-factor likelihood (solid lines) with the constraints given by the Fermi collaboration (dot lines). The red and blue lines represent the results with dark matter annihilation to $b\bar{b}$ and $\tau\bar{\tau}$ final state.}
\label{fig6}
\end{figure}
%*************************************************************************
%*************************************************************************
\begin{table}[htbp]
\centering
\caption{\label{tab:test}Listing of $J_S(\varepsilon_{\phi })$ and $J_{\rm obs}$ for the 15 dSphs.}
\begin{threeparttable}
\begin{tabular}{cccccccccccccc}
\toprule
\hline
Name & $\log(J_S(1.0))$\tnote{\it ab} & $\log(J_S(0.6))$\tnote{\it ab} & $\log(J_S(0.1))$\tnote{\it ab} & $\log(J_S(0.05))$\tnote{\it ab} & $\log(J_{\rm obs})$\tnote{\it b} \\
\midrule                                                       
Bootes I & 19.87 & 23.07 & 20.48 & 20.78 & 18.8 \\
Canes Venatici II & 18.90 & 22.10 & 19.51 & 19.81 & 17.9 \\
Carina & 19.22 & 22.42 & 19.84 & 20.13 & 18.1 \\
Coma Berenices & 20.03 & 23.23 & 20.64 & 20.94 & 19.0 \\
Draco & 19.95 & 23.15 & 20.57 & 20.86 & 18.8 \\
Fornax & 19.29 & 22.49 & 19.91 & 20.20 & 18.2 \\
Hercules & 19.12 & 22.32 & 19.74 & 20.04 & 18.1 \\
Leo II & 18.60 & 21.80 & 19.22 & 19.51 & 17.6 \\
Leo IV & 18.99 & 22.19 & 19.60 & 19.90 & 17.9 \\
Sculptor & 19.72 & 22.92 & 20.33 & 20.63 & 18.6 \\
Segue 1 & 20.61 & 23.81 & 21.23 & 21.52 & 19.5 \\
Sextans & 19.57 & 22.77 & 20.18 & 20.48 & 18.4 \\
Ursa Major II & 20.41 & 23.61 & 21.03 & 21.33 & 19.3 \\
Ursa Minor & 19.94 & 23.14 & 20.56 & 20.85 & 18.8 \\
Willman 1 & 20.17 & 23.37 & 20.78 & 21.08 & 19.1 \\
\hline
\bottomrule
\end{tabular}
\begin{tablenotes}
\footnotesize
 \item[\it a] $\alpha_{X}$ is fixed at $0.1$.
 \item[\it b] $J_S(\varepsilon_{\phi })$ and $J_{\rm obs}$ in unit $\rm GeV^{2}\;cm^{-5}$.
\end{tablenotes}
\end{threeparttable}
\end{table}
%*************************************************************************

Starting with Fermi's published upper limits on the gamma-ray flux \cite{online}, in the following, we determine the constraints on the dark matter parameters taking into account the Sommerfeld enhancement.
To determine the Sommerfeld-enchanced $J$-factor, we should know the parameters of the dark matter profile. The astrophysical parameters $(V_{\rm max},R_{V_{\rm max}})$ for the 15 dSphs are taken from Ref. \cite{Martinez2015}, in which a multilevel statistical modelling technique is used to constrain the dark matter halo properties (this technique is also adopted by the Fermi collaboration \cite{Fermi2014}). We find that the $J$-factors ($J_{\rm obs}$) calculated here are in good agreement with the values given by the Fermi collaboration \cite{Fermi2015}.
We show the calculation results of Sommerfeld-enchanced $J$-factor $J_S$ with various $\varepsilon_{\phi }$ for the 15 dSphs in table 1. From this table we can find that the targets which have the same $J_{\rm obs}$ may have different values of $J_S$, for instance, Carina and Hercules. $V_{\rm max}$ and $R_{V_{\rm max}}$ for these dSphs are not the same, which leads to a different dark matter particle velocity distribution.

Given the Sommerfeld-enchanced $J$-factor we calculate the gamma-ray flux produced by the dark matter annihilations in the dSphs and use the upper limits on the gamma-ray flux to determine 95\% confidence level limits on the dark matter parameters.
In figure 7, we show the constraints on $\varepsilon_{\phi}$ at 95\% confidence level as a function of dark matter mass. As is shown in the figure, all of the $\varepsilon_{\phi}$ parameter spaces have been excluded for the dark matter mass $m_{\chi}\lesssim 100\rm\;GeV$, since the Fermi-LAT experiment has the ability to detect the dark matter signal with a cross section below the canonical thermal relic cross section in this region. As is expected, the constraints from the dSphs search results are much more stringent than that from the subhalo observation results. 
In figure 8, we show the constraints on the dark matter annihilation cross section at 95\% confidence level as a function of the dark matter mass.
The coupling $\alpha_{X}$ is assumed to be $0.1$ here, and we notice that the constraints are independent of this parameter for dark matter mass $m_{\chi}\lesssim 10\rm\;TeV$. As is shown in the figure, our results highlight that the Sommerfeld enhancement hypothesis is discordant with the WIMPs scenario, these results are in line with the previous studies \cite{Feng2010-2, Bringmann2017}.
%*****************************fig7***************************************
\begin{figure}
\centering
\includegraphics[width=60mm,angle=0]{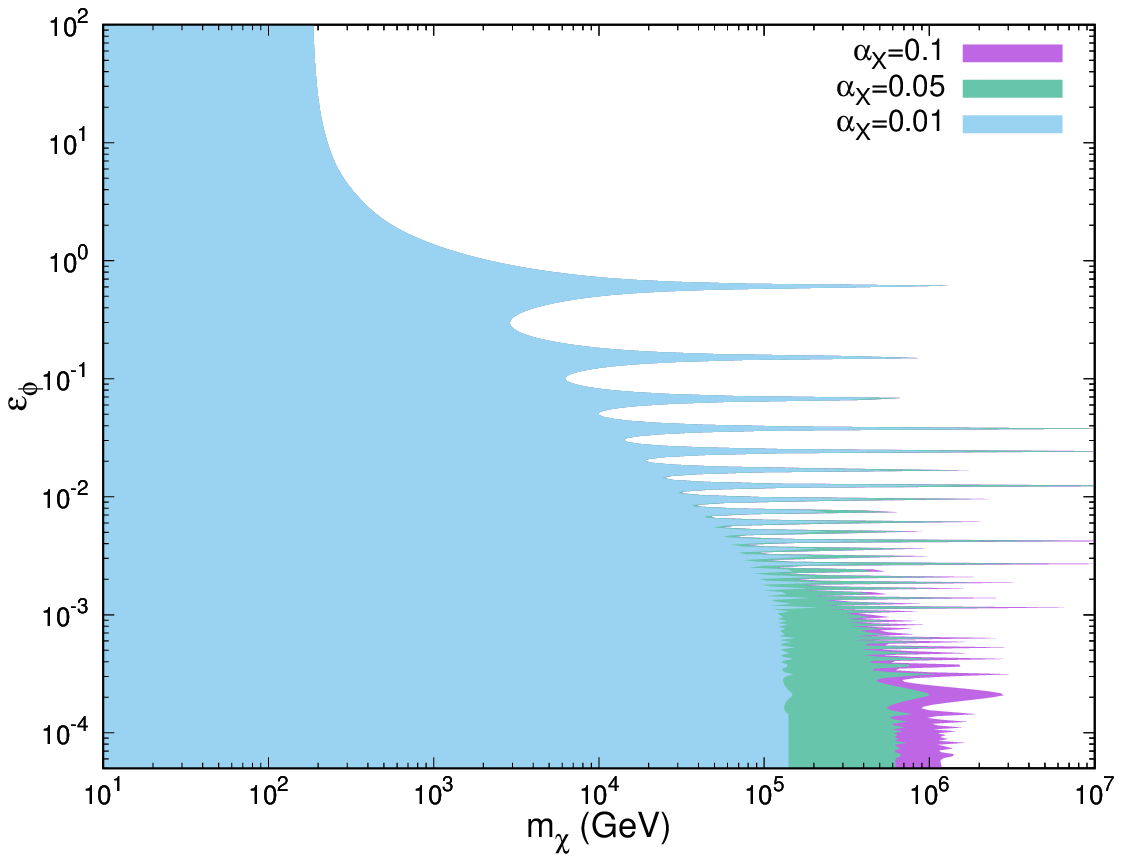}
\includegraphics[width=60mm,angle=0]{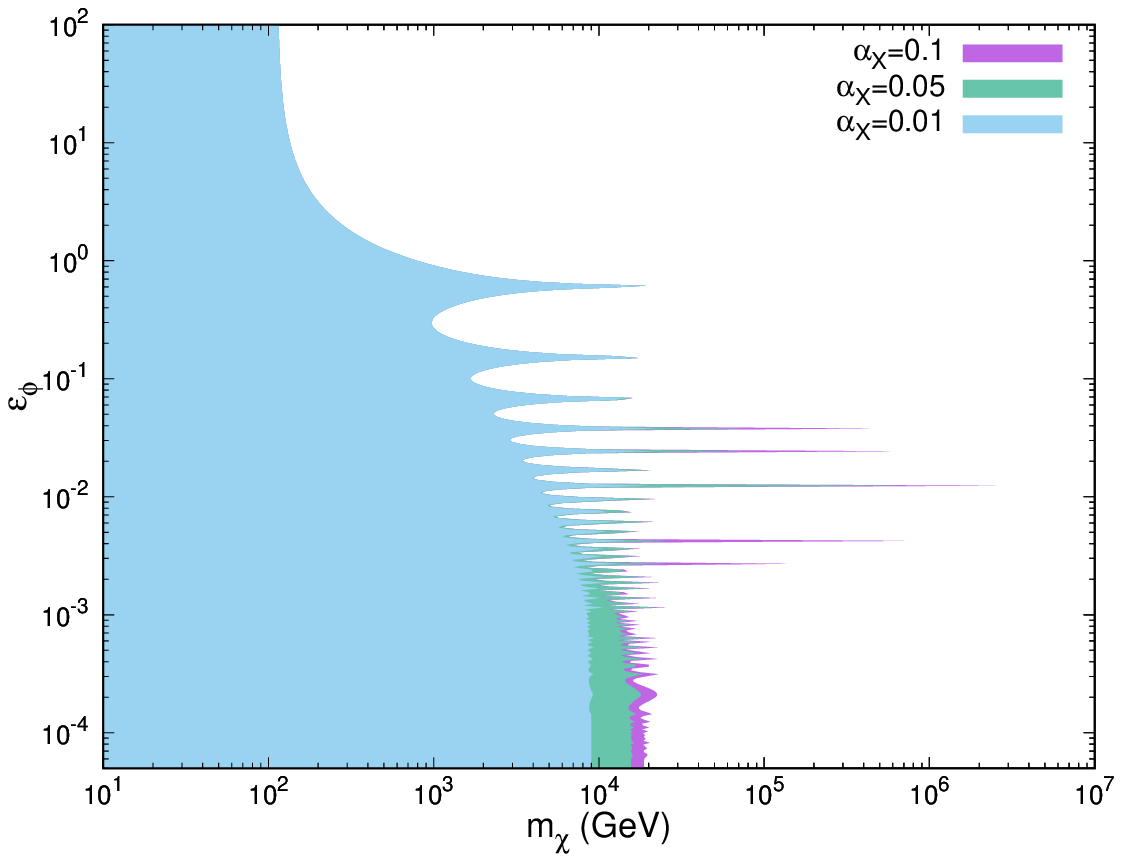}
\caption{The Fermi-LAT dSphs searches constraints on $\varepsilon_{\phi}$ at 95\% confidence level. The color shaded parameter space is excluded at 95\% confidence level, the dark matter
annihilation channel is assumed to be $b\bar{b}$ and $\tau\bar{\tau}$ in the left and right panel.}
\label{fig7}
\end{figure}

%*************************************************************************
%*****************************fig8***************************************
\begin{figure}
\centering
\includegraphics[width=60mm,angle=0]{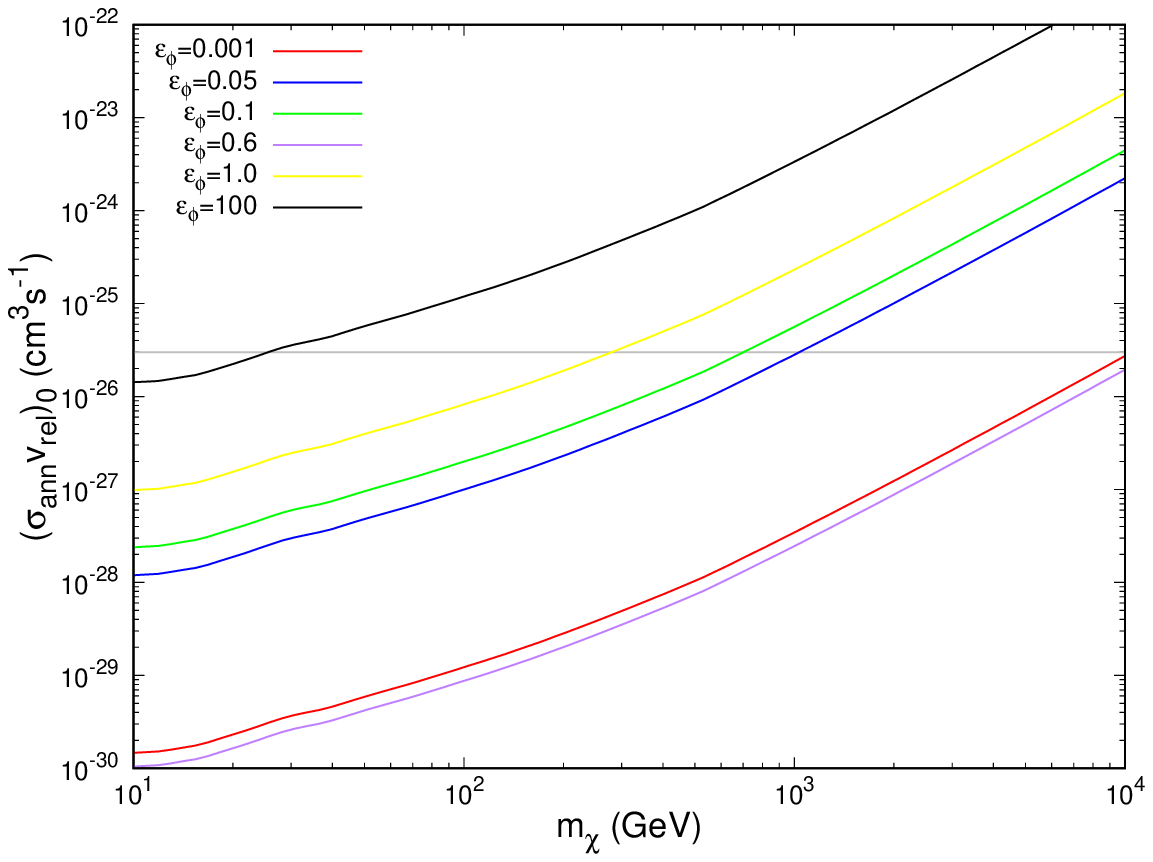}
\includegraphics[width=60mm,angle=0]{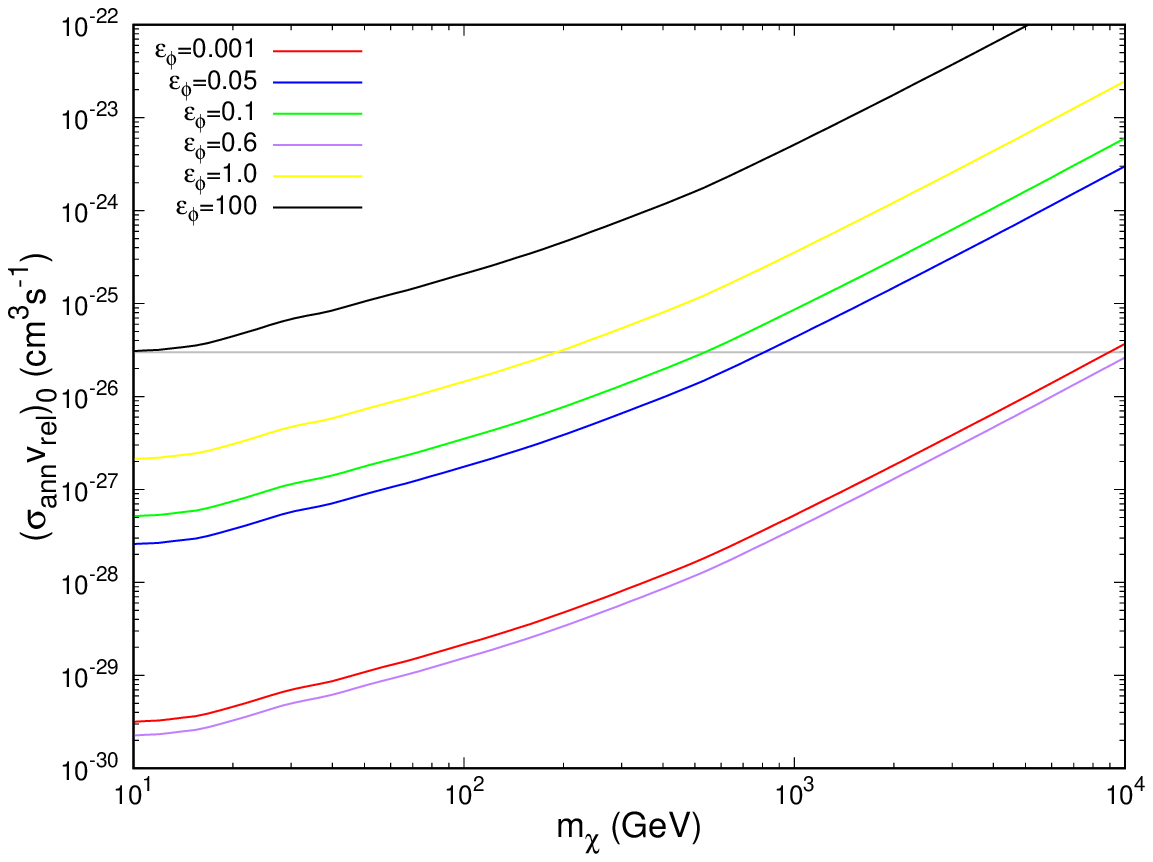}\\
\includegraphics[width=60mm,angle=0]{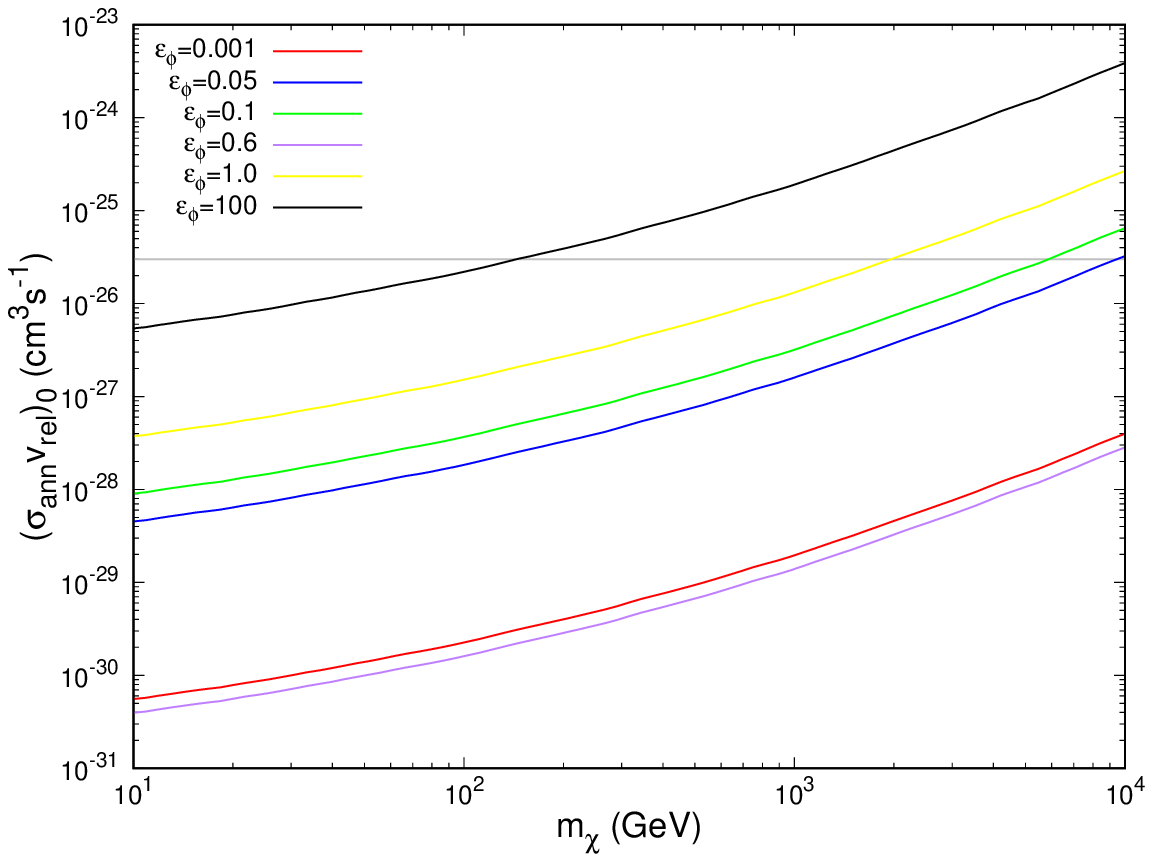}
\includegraphics[width=60mm,angle=0]{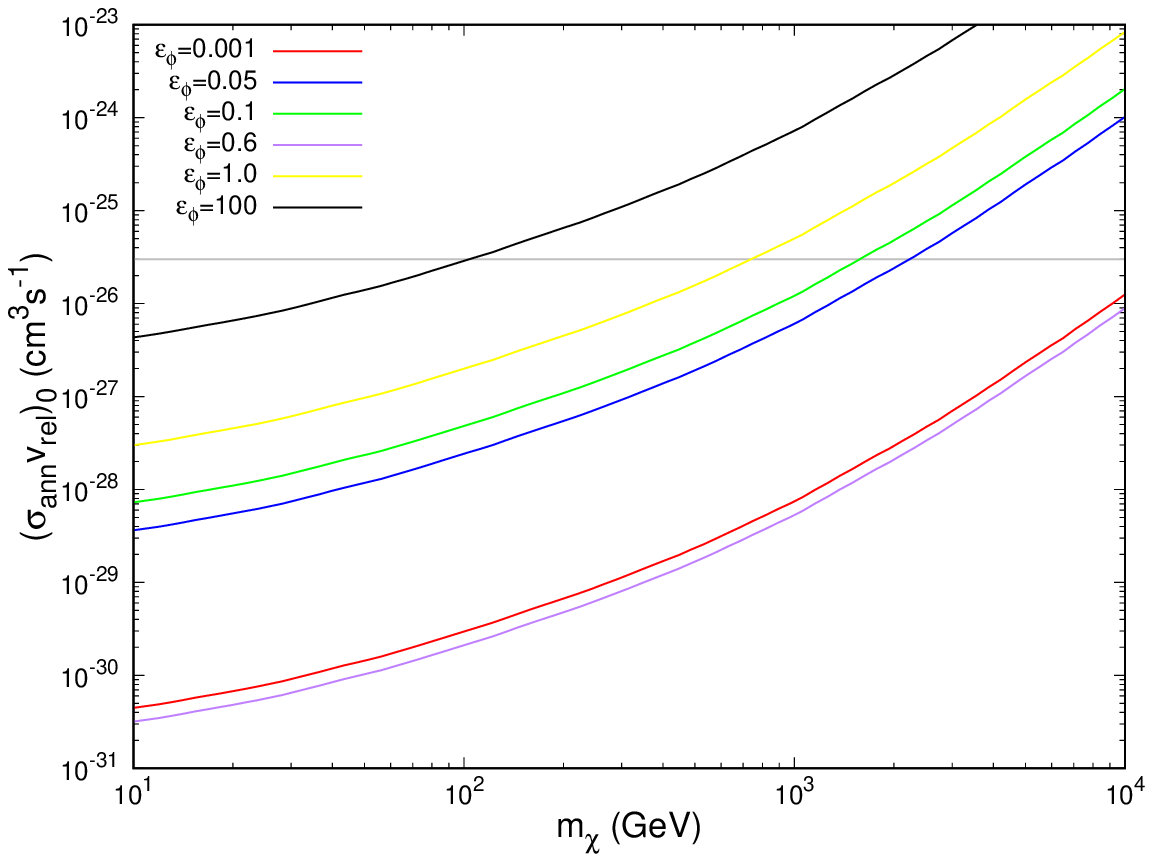}
\caption{Constraints on the dark matter annihilation cross section at 95\% confidence level derived from the Fermi-LAT dSphs searches, with a given Sommerfeld-enhanced $J$-factor. The horizontal grey line stands for an annihilation cross section $(\sigma_{\rm ann}v_{\rm rel})_0=3\times 10^{-26}\rm\;cm^{3}\;s^{-1}$, the dark matter annihilation channel is assumed to be $e^+e^-$, $\mu^+\mu^-$, $b\bar{b}$ and $\tau\bar{\tau}$ from the upper left panel to the lower right panel. }
\label{fig8}
\end{figure}

%*************************************************************************

The anomaly in the cosmic positron fraction data at energies $\gtrsim 10\rm\;GeV$ has motivated dark matter candidates with Sommerfeld-enhanced annihilation. Ref. \cite{Jin2013} gives comprehensive studies on the AMS-02 positron fraction data. They show that for the case of model A of cosmic ray propagation (see table 1 of Ref. \cite{Jin2013}) and dark matter annihilations to $b\bar{b}$, the required enhancement is in the range of $\sim (200-2000)$ and the dark matter mass is in the range of $\sim (700-2500)\rm\;GeV$ (bounds on the dark matter annihilation cross section are shown in figure 5 of Ref. \cite{Jin2013}).
In the left panel of figure 9, we show the results of Sommerfeld enhancement with various coupling $\alpha_{X}$. The dark matter particle velocity is taken to be $220\rm\;km/s$, which is a typical value for the Galactic dark matter. As is shown in the figure, to reach the required enhancement we should require a strong coupling to the light force carrier, $\alpha_{X}\gtrsim 0.1$.
We then fix the coupling $\alpha_{X}$ at 0.1 and randomly generate the parameters points with $\varepsilon_{\phi}$ in $10^{-2}-1$ and dark matter mass in the range of $(700-2500)\rm\;GeV$ (notice that as is shown above the constraints are independent of $\alpha_{X}$ for dark matter mass $m_{\chi}\lesssim 10\rm\;TeV$). We select out the points with a Sommerfeld enhancement $S>150$. Through these procedures we estimate the parameter spaces that may account for the AMS-02 positron fraction anomaly, the results are shown by the green bands in the right panel of figure 9. The constraints from the Fermi-LAT dSphs searches are represented by the red line, these results show that the Sommerfeld enhancement parameter spaces that may account for the AMS-02 positron anomaly are restricted by the Fermi-LAT gamma-ray observation results.
%*****************************fig9***************************************
\begin{figure}
\centering
\includegraphics[width=60mm,angle=0]{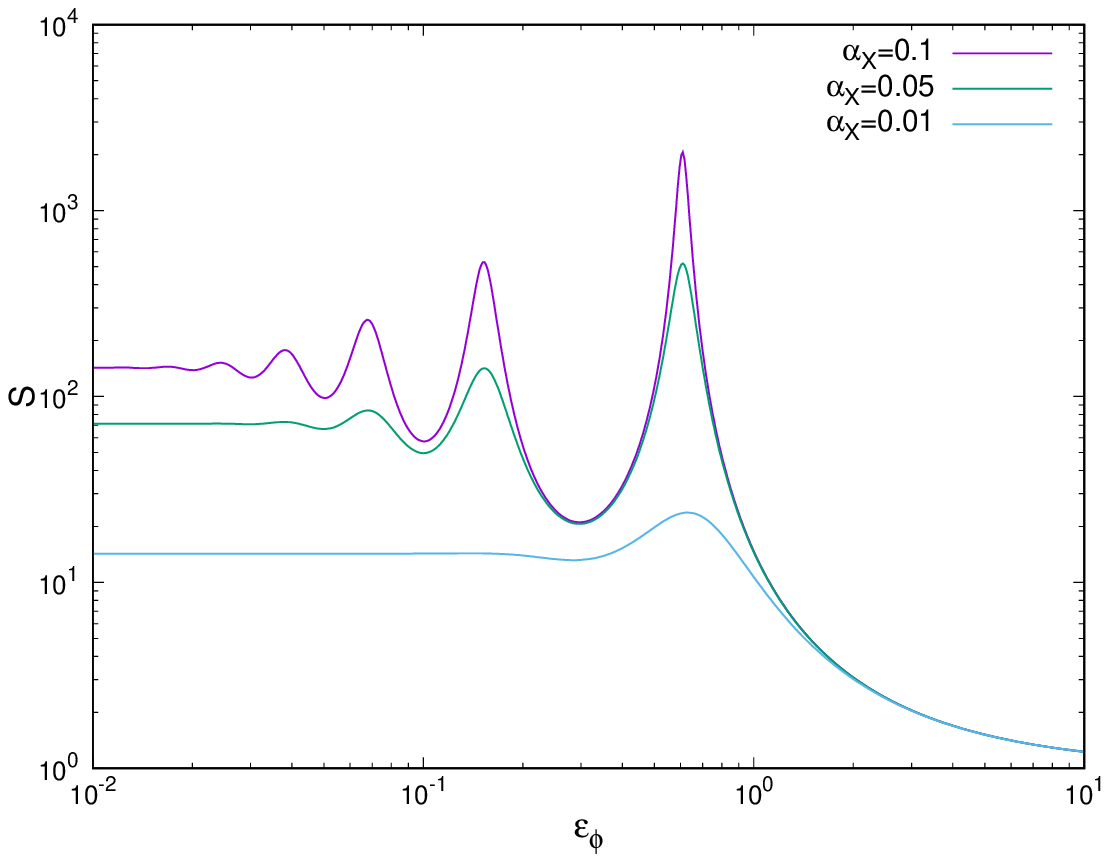}
\includegraphics[width=60mm,angle=0]{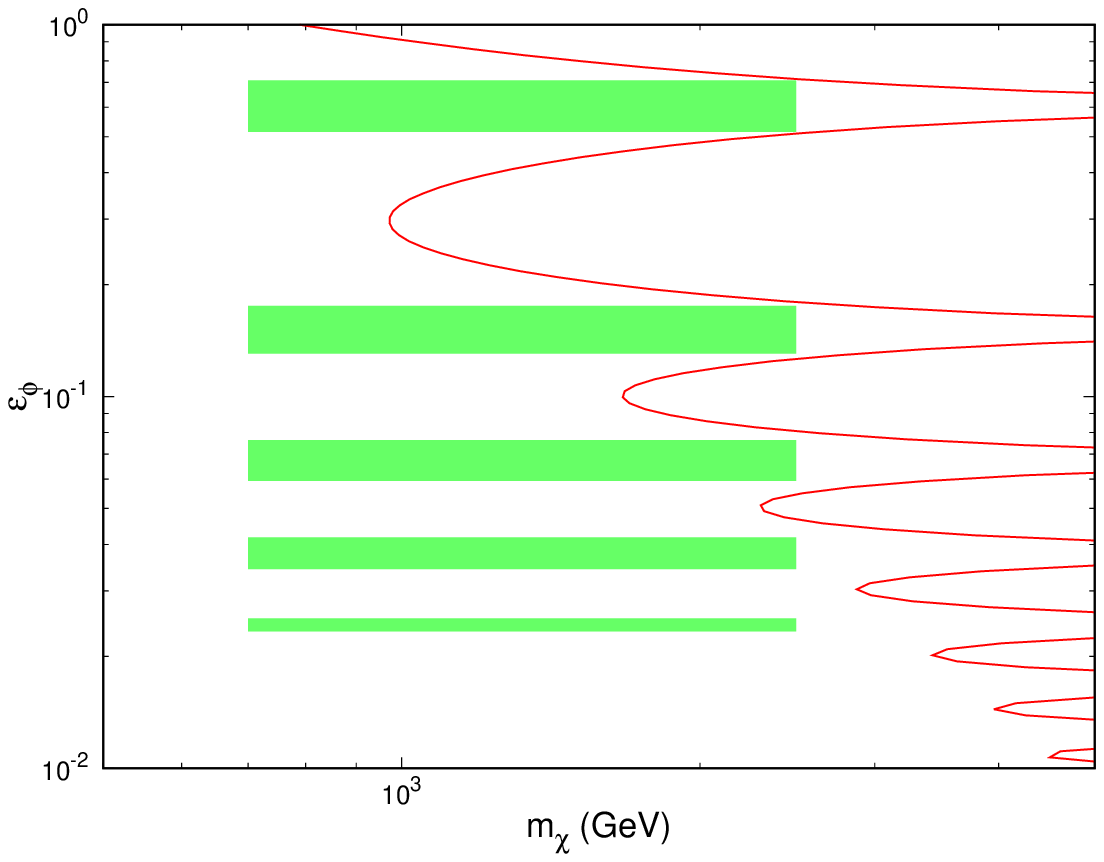}
\caption{Left panel: Plots of Sommerfeld enhancement as a function of $\varepsilon_{\phi}$ with various $\alpha_{X}$, assuming dark matter particle velocity $v=220\rm\;km/s$. Right panel: the green bands stand for the parameter points with the Sommerfeld enhancement $S>150$, the red line represents the constraints from the Fermi-LAT dSphs observations.}
\label{fig9}
\end{figure}

%*************************************************************************

\section{Conclusions}

In the case of velocity-independent dark matter cross section, $J$-factor is independent of the underlying dark matter particle physics and can be simply factorized from the particle physics.
However, the annihilation cross section cannot be extracted from the $J$-factor directly when the annihilation cross section is velocity-dependent, since the photon flux arising from dark matter annihilation depends on the dark matter velocity distribution.
In this work we focus on the dark matter annihilation cross section that is enhanced by the Sommerfeld effect. We determine the dark matter velocity distribution for the NFW dark matter density profile using the Eddington's formula, with the assumption that the orbits of the dark matter particles are isotropic. We define a dimensionless dark matter velocity distribution function and show that this distribution function can be well fitted by an exponential form of velocity. We then calculate the Sommerfeld-enhanced $J$-factor with the dark matter velocity distribution function and show its behaviors in many parameter spaces.

The subhalos and dSphs have low characteristic dark matter particle velocities, thus are ideal sites to study the Sommerfeld enhancement effect. We use the Fermi-LAT gamma-ray sources observation results to put constraints on the Sommerfeld enhancement.
For the subhalo observations, we count the predicted numbers of dark matter sources that may be observed by the Fermi-LAT experiment and determine the 95\% confidence level Poisson upper limit on the dark matter annihilation cross section. For the dSphs, we use the likelihood and upper limits on the gamma-ray flux provided by the Fermi collaboration to determine the upper limits on the dark matter parameters space at 95\% confidence level.
Fitting to the AMS-02 positron fraction data shows that to account for the positron anomaly with dark matter scenario, the required enhancement and dark matter mass should be in the range of $(200-2000)$ and $(700-2500)\rm\;GeV$. Our analysis shows that these parameter spaces have been restricted by the Fermi-LAT gamma-ray observations.
Our results may also put stringent constraints on the self-interacting dark matter models that may account for the Galactic Center GeV gamma-ray excess.

Blum et al. \cite{Blum2016} have shown that the naive Sommerfeld enhancement $S$ used in this work should be replaced by the regulated Sommerfeld enhancement $\tilde{S}$ when the partial-wave unitarity bounds are taken into account (see the original work \cite{Blum2016} for more details).
Depending on the choice of parameters (for instance, the coupling $\alpha_{X}$, mediator mass, and short-range annihilation and elastic cross sections), the enhancement close to a resonant point may be many orders of magnitude smaller in the regulated case.
Thus, the constraints near resonance peaks could be weaker than that evaluated in this work if we use the regulated Sommerfeld enhancement. 
Notice that figure 2 of Ref. \cite{Blum2016} shows that the discrepancy between the naive and regulated Sommerfeld enhancement near a resonance peak goes to zero when $v\gtrsim 10^{-4}$, assuming specific values of the parameters. Further research should be done on this issue for the dwarf galaxy system, in which the velocity of dark matter particle is in the range $10^{-5}-10^{-4}$.

\acknowledgments{This work is supported in part by the National Key Research and Development Program of China 
No. 2017YFA0402200 and 2017YFA0402204, the NSFC under Grants No. 11335012 and No. 11475237, and the Key Research Program of Frontier Sciences, CAS.
}

\appendix
\section{Dimensionless velocity distribution function}

The NFW density profile can be expressed in the scaled form
\begin{eqnarray}
\rho(r)=\rho_s\times \tilde{\rho }(\tilde{r}),
\end{eqnarray}
where $\tilde{\rho }$ is a dimensionless quantity as a function of $\tilde{r}=r/r_s$. Then the gravitational potential can be written as 
\begin{eqnarray}
\Psi (r)=G\rho_{s}r_{s}^{2}\times \tilde{\Psi }(\tilde{r}),
\end{eqnarray}
where the dimensionless gravitational potential is given by
\begin{eqnarray}
\tilde{\Psi }(\tilde{r})=\begin{cases}
 -C/\tilde{r}_{\rm max}-4\pi\int_{\tilde{r}}^{\tilde{r}_{\rm max}}(\ln(1+\tilde{r})-\tilde{r}/(1+\tilde{r}))/\tilde{r}^{2}d\tilde{r} & \text{ if } \tilde{r}< \tilde{r}_{\rm max}   \\
 -C/\tilde{r} & \text{ if } \tilde{r}\ge \tilde{r}_{\rm max}  
\end{cases}
\end{eqnarray}
where $C=4\pi(\ln(1+\tilde{r}_{\rm max})-\tilde{r}_{\rm max}/(1+\tilde{r}_{\rm max}))$ and $\tilde{r}_{\rm max}=2.163$.
We can also define a dimensionless energy per unity mass 
\begin{eqnarray} 
\tilde{\varepsilon }=\tilde{v}^{2}/2+\tilde{\Psi }(\tilde{r}),
\end{eqnarray}
where $\tilde{v}=(G\rho_{s}r_{s}^{2})^{-1/2}v$ is a dimensionless velocity. With these dimensionless quantities, the dark matter distribution function can be written as
\begin{eqnarray} 
f(\varepsilon )=\rho_{s}(G\rho_{s}r_{s}^2)^{-3/2}\tilde{f}(\tilde{\varepsilon }),
\end{eqnarray}
where $\tilde{f}(\tilde{\varepsilon })$ is the dimensionless dark matter velocity distribution function, which has the same form as Eq. (2.1)
\begin{eqnarray} 
\tilde{f}(\tilde{\varepsilon })=\frac{1}{\sqrt{8}\pi^2}\int_{\tilde{\varepsilon }}^{0}\frac{d^{2}\tilde{\rho }}{d\tilde{\Psi }^2}\frac{d\tilde{\Psi }}{\sqrt{\tilde{\Psi}-\tilde{\varepsilon }}}.
\end{eqnarray}
We find this dimensionless dark matter velocity distribution function can be parameterized as
\begin{eqnarray} 
\tilde{f}(\tilde{r},\tilde{v})=\exp\left ( \sum_{n=0}^{\infty }a_{n}(\tilde{r})\tilde{v}^{n} \right ),
\end{eqnarray}
where $a_{n}(\tilde{r})$ is a coefficient that depends on the dimensionless radius $\tilde{r}$. Expanding the series to $n=5$ or $n=6$ can give well fitting results with this formula.
We show the numerical and fitting results of dimensionless distribution function $\tilde{f}(\tilde{r},\tilde{v})$ in figure 10. The purple lines in the figure represent the numerical results and the green lines stand for the fitting results, from the upper left panel to the lower right panel the ``radius" $\tilde{r}$ are set at 0.1, 0.5, 1.0 and 1.5 respectively. As is shown in the figure, the dimensionless distribution function $\tilde{f}(\tilde{r},\tilde{v})$ is nearly a constant when  ``velocity" $\tilde{v}\lesssim 0.1$, while it drops rapidly to zero as ``velocity" increases. The figure also shows that dimensionless distribution function $\tilde{f}(\tilde{r},\tilde{v})$ decreases with the increasing of ``radius". 
%*****************************fig10***************************************
\begin{figure}
\centering
\includegraphics[width=60mm,angle=0]{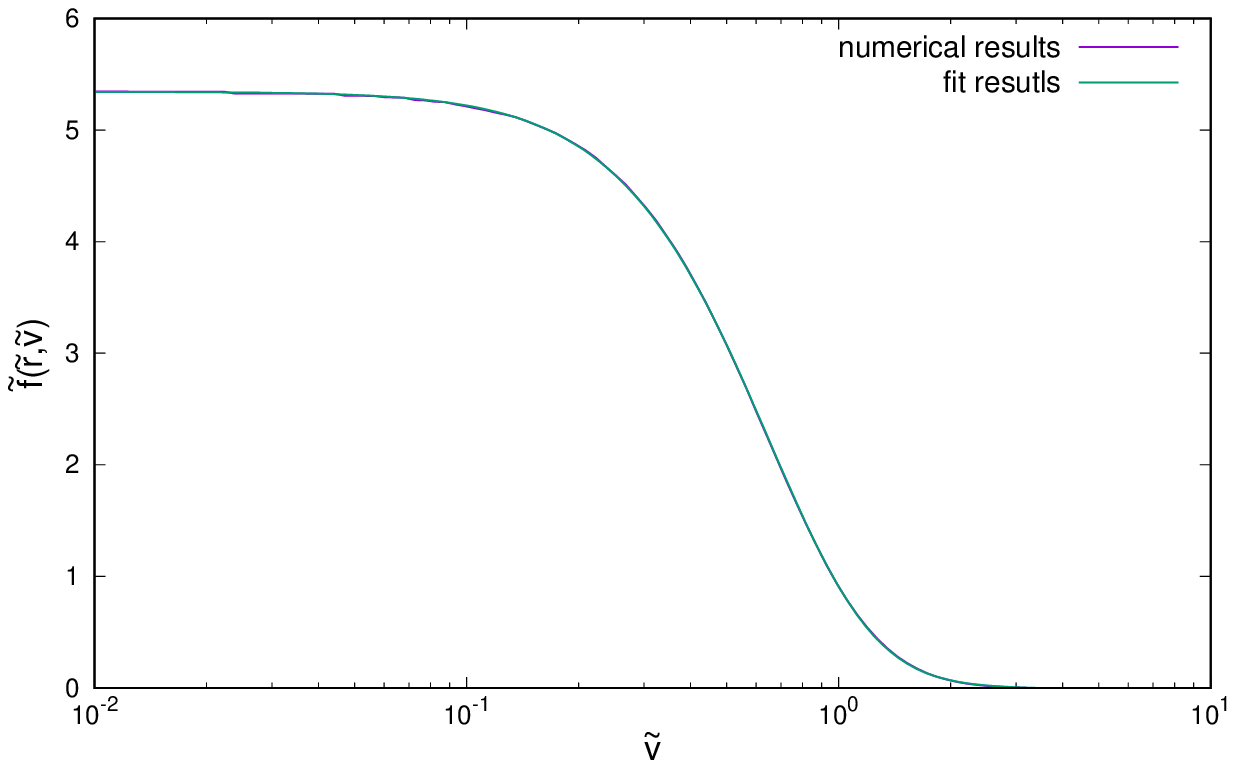}\includegraphics[width=60mm,angle=0]{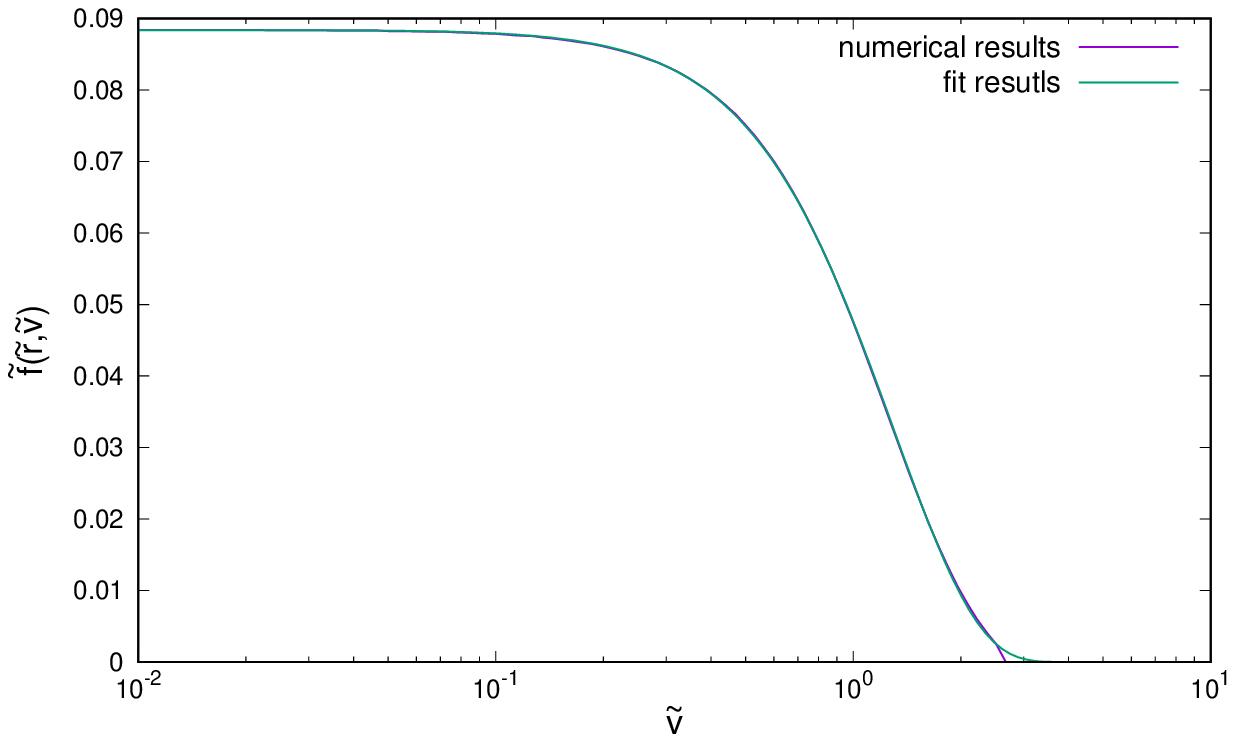}\\
\includegraphics[width=60mm,angle=0]{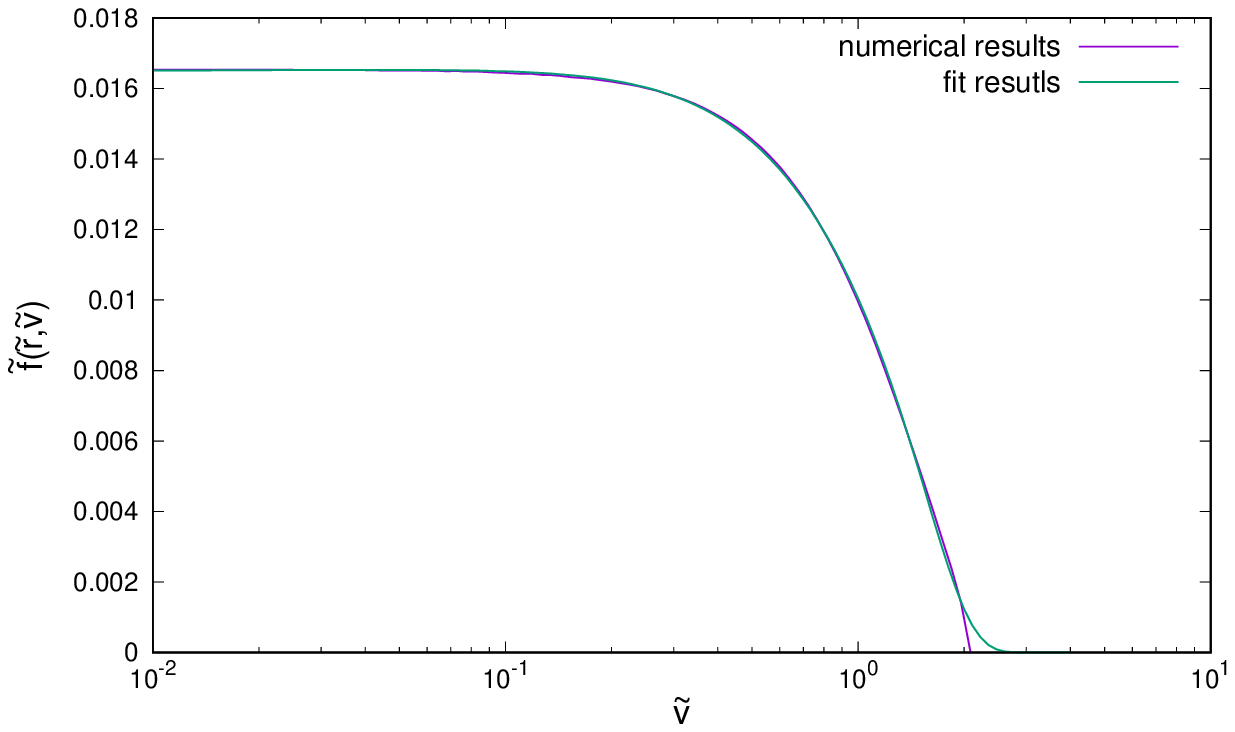}\includegraphics[width=60mm,angle=0]{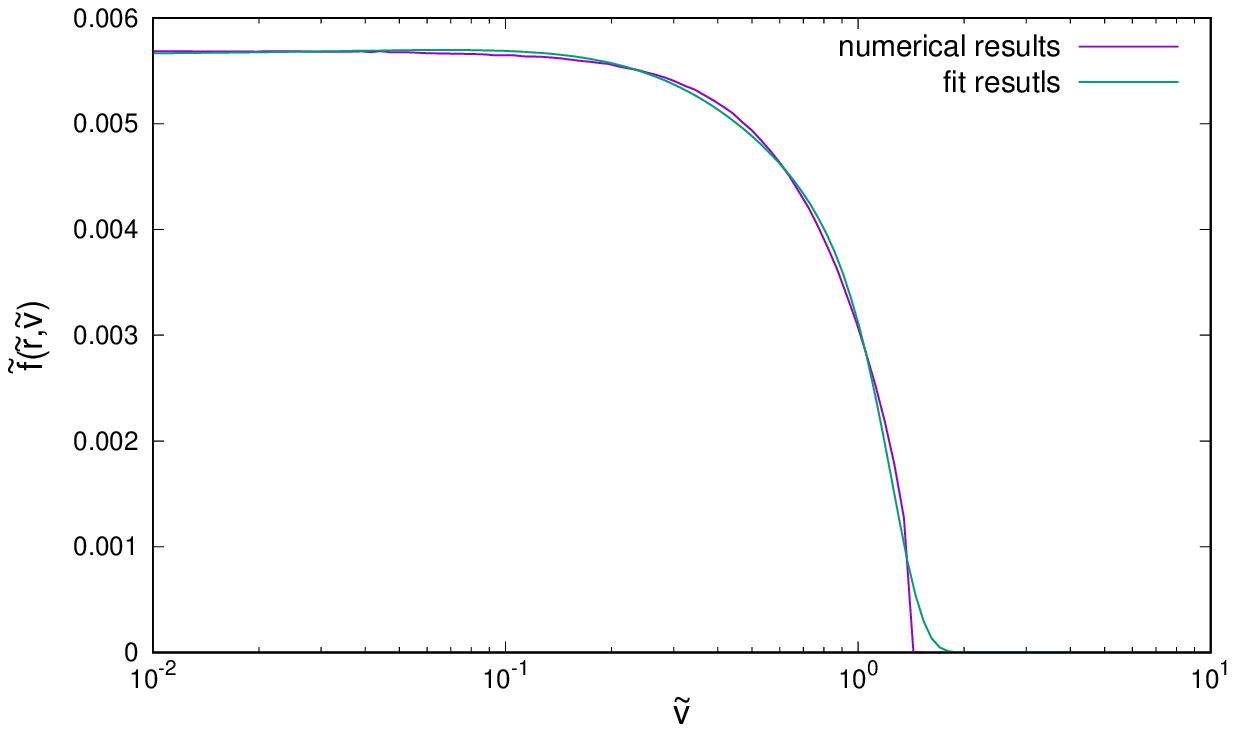}
\caption{Numerical and fitting results of dimensionless velocity distribution function $\tilde{f}(\tilde{r},\tilde{v})$ as a function $\tilde{r}$. The ``radius" $\tilde{r}$ is set at 0.1 and 0.5 for the upper two panels, and for the lower two panels the ``radius" $\tilde{r}$ is set at 1.0 and 1.5 respectively.}
\label{fig10}
\end{figure}

%*************************************************************************

\section{Analytic solutions for the Sommerfeld enhancement}
In the limit of a massless force carrier, the Yukawa potential becomes a Coulomb potential and can be solved analytically \cite{Sommerfeld1931}. In this case the Sommerfeld enhancement factor is given by
\begin{eqnarray}
S=\frac{\pi/\varepsilon_{v}}{1-e^{-\pi/\varepsilon_{v}}}.
\end{eqnarray}
In the low-velocity regime $v\ll \alpha_{X}$, $S\simeq \pi\alpha_{X}/v$.
There are no resonances in the Sommerfeld enhancement for the Coulomb case, because the potential is not localized.
We notice that the behavior of Sommerfeld enhancement decreasing with $1/v$ breaks down for very small velocities \cite{Lattanzi2009}. 
To see this, we can expand the Yukawa potential in powers of $y=m_{\phi}r$
\begin{eqnarray}
V(r)\simeq -\frac{\alpha_{X}}{r}(1-m_{\phi}r)=V_{\rm c}(r)(1-m_{\phi}r),
\end{eqnarray}
where $V_{\rm c}(r)$ is the Coulomb potential. The radial Schr\"odinger equation becomes
\begin{eqnarray}
\frac{1}{m_{\chi}}{\psi}''(r)-V_{\rm c}(r)\psi(r)=-(m_{\chi}v^{2}-m_{\phi}\alpha)\psi(r).
\end{eqnarray}
The Coulomb case can be recovered by the Yukawa case when the condition $v^2\gg m_{\phi}\alpha_{X}/m_{\chi}$ holds.
The Sommerfeld enhancement saturates at low velocity for a nonzero mass of mediator $\phi$. Since the attractive force has a finite range $\sim 1/m_{\phi}$, the Sommerfeld enhancement saturates at $S\sim 1/\varepsilon_{\phi}$ when the deBroglie wavelength of the particle $1/(m_{\chi}v)$ gets larger than the interaction range \cite{Hamed2009}.
%*****************************fig11***************************************
\begin{figure}
\centering
\includegraphics[width=60mm,angle=0]{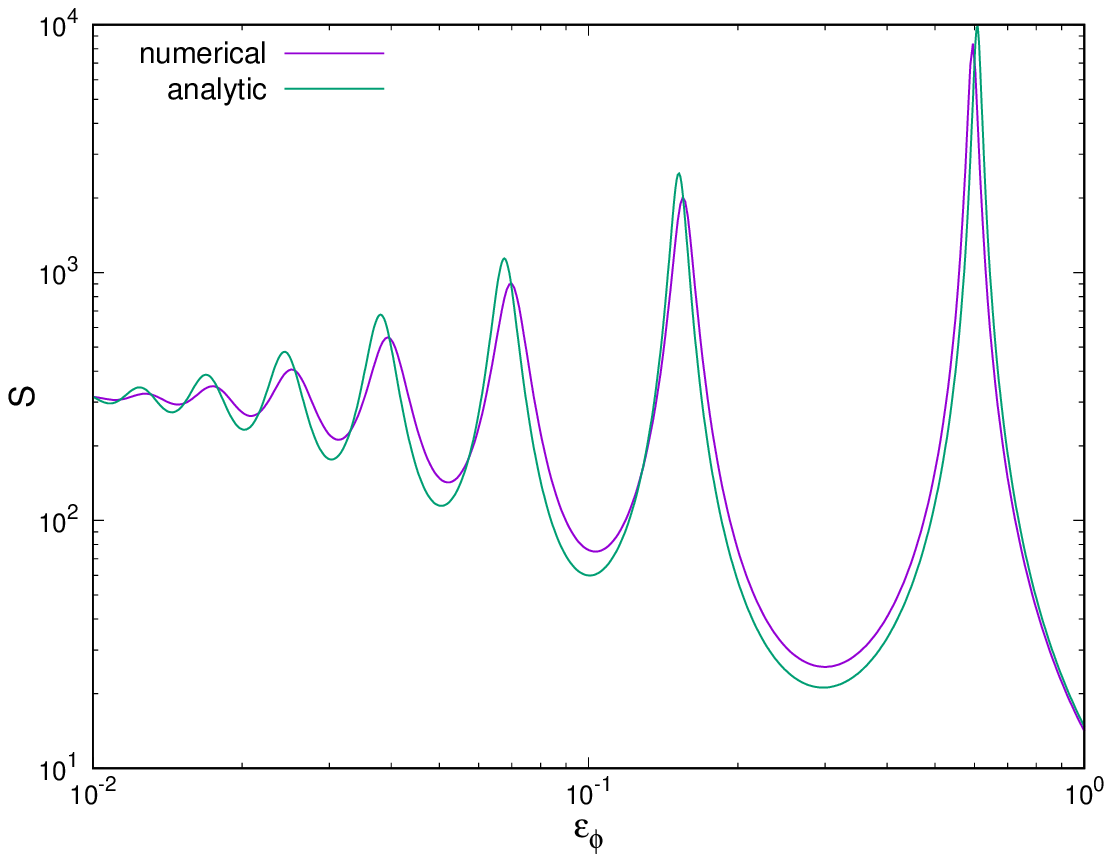}
\includegraphics[width=60mm,angle=0]{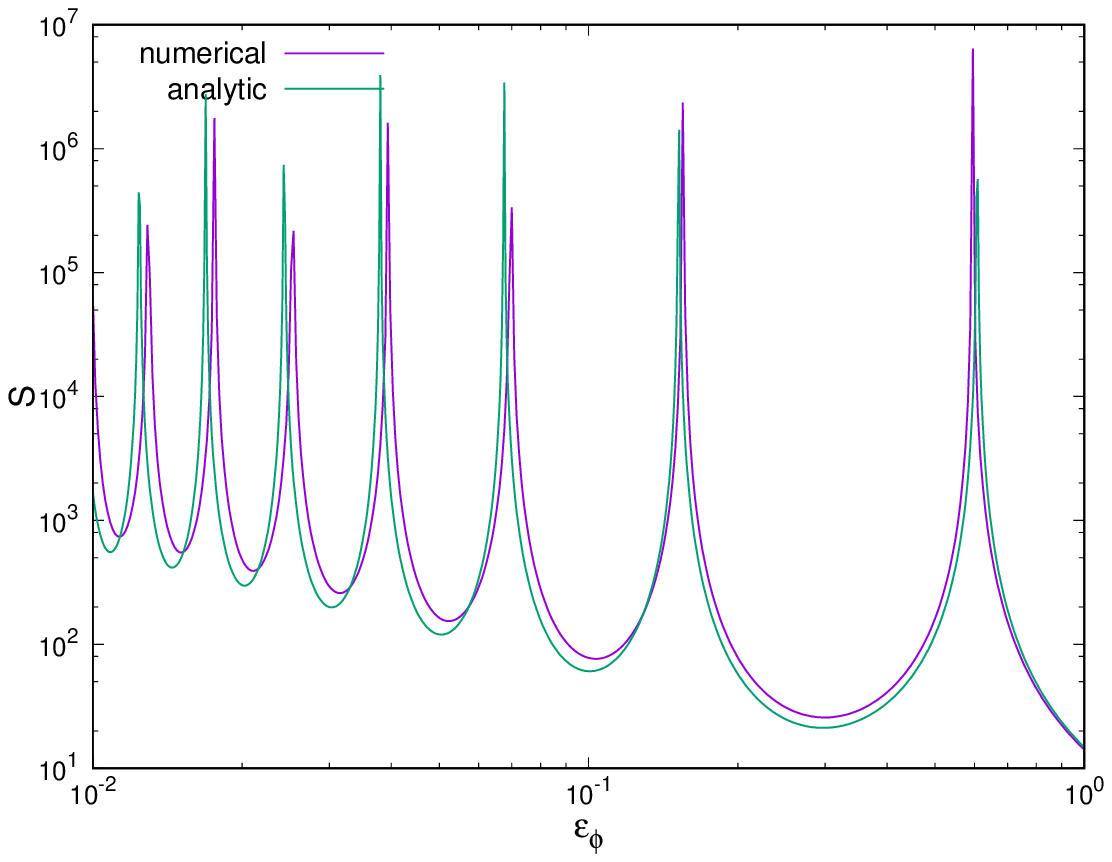}
\caption{Numerical and analytic solutions of the Sommerfeld enhancement factor $S$ as a function of $\varepsilon_{\phi}$, evaluated at $\varepsilon_{v}=10^{-2}$ (left panel) and $\varepsilon_{v}=10^{-4}$ (right panel) respectively.}
\label{fig11}
\end{figure}

%*************************************************************************
The analytic solution for the Schr\"odinger equation can also be found by approximating the Yukawa potential as the Hulth$\acute{\rm e}$n potential \cite{Cassel2010, Slatyer2010}, and the resulting Sommerfeld enhancement is given by
\begin{eqnarray}
S=\frac{\pi}{\varepsilon_{v}}\frac{\sinh\left ( \frac{2\pi\varepsilon_{v}}{\pi^{2}\varepsilon_{\phi}/6} \right )}{\cosh\left ( \frac{2\pi\varepsilon_{v}}{\pi^{2}\varepsilon_{\phi}/6} \right )-\cos\left ( 2\pi\sqrt{\frac{1}{\pi^{2}\varepsilon_{\phi}/6}-\frac{\varepsilon_{v}^{2}}{(\pi^{2}\varepsilon_{\phi}/6)^{2}}} \right )}.
\end{eqnarray}
The analytic result is an excellent approximation, typically reproducing the numerical results to within fractional differences of 10\%, while in the resonance regions, the discrepancies may be as large as a factor of 2 or beyond \cite{Feng2010-2, Boddy2017}.
We show the analytic approximation (green lines) and numerical results (purple lines) of the Sommerfeld enhancement in figure 11.


\begin{thebibliography}{}

\bibitem{Jungman1996} G. Jungman, M. Kamionkowski and K. Griest, {\it Supersymmetric dark matter, Phys. Rept.} {\bf 267} (1996) 195 [hep-ph/9506380].

\bibitem{Bergstrom2000} L. Bergstr\"om, {\it Nonbaryonic dark matter: observational evidence and detection methods, Rept. Prog. Phys.} {\bf 63} (2000) 793 [hep-ph/0002126].

\bibitem{Bertone2005RP} G. Bertone, D. Hooper and J. Silk, {\it Particle dark matter: evidence, candidates and constraints, Phys. Rept.} {\bf 405} (2005) 279 [hep-ph/0404175].

\bibitem{Lee1977} B. W. Lee and S. Weinberg, {\it Cosmological lower bound on heavy-neutrino masses Phys. Rev. Lett.} {\bf 39} (1977) 165.

\bibitem{Hut1977} P. Hut, {\it Limits on masses and number of neutral weakly interacting particles Phys. Lett.} {\bf 69B} (1977) 85.

\bibitem{Merritt2002} D. Merritt, M. Milosavljevic, L. Verde and R. Jimenez, {\it Dark matter spikes and annihilation radiation from the galactic center, Phys. Rev. Lett.} {\bf 88} (2002) 191301 [astro-ph/0201376].

\bibitem{Cesarini2004} A. Cesarini, F. Fucito, A. Lionetto, A. Morselli and P. Ullio, {\it The Galactic Center as a dark matter gamma-ray source, Astropart. Phys.} {\bf 21} (2004) 267 [astro-ph/0305075].

\bibitem{Dodelson2008} S. Dodelson, D. Hooper and P. D. Serpico, {\it Extracting the gamma ray signal from dark matter annihilation in the galactic center region, Phys. Rev. D} {\bf 77} (2008) 063512 [arXiv:0711.4621].

\bibitem{Stoehr2003} F. Stoehr et al., {\it Dark matter annihilation in the halo of the Milky Way, Mon. Not. Roy. Astron. Soc.} {\bf 345} (2003) 1313 [astro-ph/0307026].

\bibitem{Bergstrom1998} L. Bergstr\"om, P. Ullio and J. H. Buckley, {\it Observability of gamma-rays from dark matter neutralino annihilations in the Milky Way halo, Astropart. Phys.} {\bf 9} (1998) 137 [astro-ph/9712318].

\bibitem{Bertone2005PRD} G. Bertone, A. R. Zentner and J. Silk, {\it A new signature of dark matter annihilations: gamma-rays from intermediate-mass black holes, Phys. Rev. D} {\bf 72} (2005) 103517 [astro-ph/0509565].

\bibitem{Brun2007} P. Brun, G. Bertone, J. Lavalle, P. Salati and R. Taillet, {\it Antiproton and positron signal enhancement in dark matter mini-spikes scenarios, Phys. Rev. D} {\bf 76} (2007) 083506 [arXiv:0704.2543].

\bibitem{Bringmann2009} T. Bringmann, J. Lavalle and P. Salati, {\it Intermediate mass black holes and nearby dark matter point sources: a myth-buster, Phys. Rev. Lett.} {\bf 103} (2009) 161301 [arXiv:0902.3665].

\bibitem{Bringmann2012} T. Bringmann and C. Weniger, {\it Gamma ray signals from dark matter: concepts, status and prospects, Phys. Dark Univ.} {\bf 1} (2012) 194 [arXiv:1208.5481].

\bibitem{Klasen2015} M. Klasen, M. Pohl and G. Sigl, {\it Indirect and direct search for dark matter, Prog. Part. Nucl. Phys.} {\bf 85} (2015) 1 [arXiv:1507.03800].

\bibitem{Frenk2012} C. S. Frenk and S. D. M. White, {\it Dark matter and cosmic structure, Annalen Phys.} {\bf 524} (2012) 507 [arXiv:1210.0544].

\bibitem{White1991} S. D. M. White and C. S. Frenk, {\it Galaxy formation through hierarchical clustering, Astrophys. J.} {\bf 379} (1991) 52.

%\bibitem{Frenk2012} C. S. Frenk and S. D. M. White, {\it Dark matter and cosmic structure, Annalen Phys.} {\bf 524} (2012) 507 [arXiv:1210.0544].

\bibitem{Moline2017} $\acute{\rm A}.$ Molin$\acute{\rm e}$, M. A. S$\acute{\rm a}$nchez-Conde, S. Palomares-Ruiz and F. Prada, {\it Characterization of subhalo structural properties and implications for dark matter annihilation signals, Mon. Not. R. Astron. Soc.} {\bf 466} (2017) 4 [arXiv:1603.04057].

\bibitem{Ackermann2012} M. Ackermann et al., {\it Search for dark matter satellites using Fermi-LAT, Astrophys. J.} {\bf 747} (2012) 121 [arXiv:1201.2691].

%\bibitem{Boehm2004-1} C. Boehm, T. Ensslin and J. Silk, {\it Can Annihilating dark matter be lighter than a few GeVs?, J. Phys. G} {\bf 30} (2004) 279 [astro-ph/0208458].

\bibitem{Diamanti2014} R. Diamanti, L. L. Honorez, O. Mena, S. P. Ruiz and A. C. Vincent, {\it Constraining dark matter late-time energy injection: decays and $p$-wave annihilations, J. Cosmol. Astropart. Phys.} {\bf 02} (2014) (017) [arXiv:1308.2578].

\bibitem{Boehm2004-2} C. Boehm and P. Fayet, {\it Scalar dark matter candidates, Nucl. Phys. B} {\bf 683} (2004) 219 [hep-ph/0305261].

%\bibitem{Boehm2004-3} C Boehm, D Hooper, J Silk, M Casse, J Paul, MeV dark matter: Has it been detected?, Phys. Rev. Lett. 92 (2004) 101301 [arXiv:astro-ph/0309686].

\bibitem{Robertson2009} B. E. Robertson and A. R. Zentner, {\it Dark matter annihilation rates with velocity dependent annihilation cross sections, Phys. Rev. D} {\bf 79} (2009) 083525 [arXiv:0902.0362].

\bibitem{Alves2014} A. Alves, S. Profumo, F. S. Queiroz and W. Shepherd, {\it Effective field theory approach to the Galactic Center gamma-ray excess, Phys. Rev. D} {\bf 90} 115003 (2014).

\bibitem{Berlin2014-1} A. Berlin, D. Hooper and S. D. McDermott, {\it Simplified dark matter models for the Galactic Center gamma-ray excess, Phys. Rev. D} {\bf 89} (2014) 115022 [arXiv:1404.0022].

%\bibitem{Profumo2013} S. Profumo and W. Shepherd, {\it Pitfalls of dark matter crossing symmetries Phys. Rev. D} {\bf 88}, 056018 (2013) [arXiv:1307.6277].

\bibitem{Sommerfeld1931} A. Sommerfeld, Ann. Phys. (Leipzig) {\bf 403} (1931) 257.

\bibitem{Hisano2004} J. Hisano, M. Nagai, M. Nojiri and M. Senami, {\it Explosive dark matter annihilation, Phys. Rev. Lett.} {\bf 92} (2004) 031303 [arXiv:hep-ph/0307216].

\bibitem{Hisano2005} J. Hisano, S. Matsumoto, M. Nojiri and S. Saito, {\it Nonperturbative effect on dark matter annihilation and gamma ray signature from the galactic center, Phys. Rev. D} {\bf 71}, 063528 (2005) [arXiv:hep-ph/0412403].

\bibitem{Profumo2005} S. Profumo, {\it TeV $\gamma$-rays and the largest masses and annihilation cross sections of neutralino dark matter, Phys. Rev. D} {\bf 72} (2005) 103521 [arXiv:astro-ph/0508628].

\bibitem{Cirelli2007} M. Cirelli, A. Strumia and M. Tamburini, {\it Cosmology and astrophysics of minimal dark matter, Nucl. Phys.} {\bf B787}, 152 (2007) [arXiv:0706.4071 [hep-ph]]. 

\bibitem{Hamed2009} N. A. Hamed, D. P. Finkbeiner, T. R. Slatyer and N. Weiner, {\it A Theory of Dark Matter, Phys. Rev. D} {\bf 79} (2009) 015014 [arXiv:0810.0713].

\bibitem{Feng2010-1} J. L. Feng, M. Kaplinghat and H. B. Yu, {\it Halo-Shape and Relic-Density Exclusions of Sommerfeld-Enhanced Dark Matter Explanations of cosmic-ray Excesses, Phys. Rev. Lett.} {\bf 104} (2010) 151301 [arXiv:0911.0422].

\bibitem{Feng2010-2} J. L. Feng, M. Kaplinghat and H. B. Yu, {\it Sommerfeld enhancements for thermal relic dark matter, Phys. Rev. D} {\bf 82} (2010) 083525 [arXiv:1005.4678].

\bibitem{Zavala2010} J. Zavala, M. Vogelsberger, S. D. M. White, {\it Relic density and CMB constraints on dark matter annihilation with Sommerfeld enhancement, Phys. Rev. D} {\bf 81} (2010) 083502 [arXiv:0910.5221].

\bibitem{Zhou2013PRD} Z. P. Liu, Y. L. Wu and Y. F. Zhou, {\it Sommerfeld enhancements with vector, scalar, and pseudoscalar force carriers, Phys.Rev. D} {\bf 88} (2013) 096008 [arXiv:1305.5438].

\bibitem{Zhou2013JCAP} J. Chen and Y. F. Zhou, {\it The 130 GeV gamma-ray line and Sommerfeld enhancements, J. Cosm. Astropart. Phys.} {\bf 1304} (2013) 017 [arXiv:1301.5778].

\bibitem{Das2017} A. Das, B. Dasgupta, {\it Selection Rule for Enhanced Dark Matter Annihilation, Phys. Rev. Lett.} {\bf 118} (2017) 251101 [arXiv:1611.04606].

%\bibitem{Pospelov2009} M. Pospelov and A. Ritz, {\it Astrophysical signatures of secluded dark matter, Phys. Lett.} {\bf B671}, 391 (2009).

%\bibitem{Russell2008} J. M. Russell, S. M. West, D. Cumberbatch and D. Hooper, {\it Heavy dark matter through the Higgs portal, J. High Energy Phys.} {\bf 07} (2008) 058.

\bibitem{PAMELA2009} PAMELA collaboration, O. Adriani et al., {\it An anomalous positron abundance in cosmic-rays with energies 1.5-100 GeV, Nature} {\bf 458} (2009) 607 [arXiv:0810.4995].

\bibitem{AMS2013} AMS collaboration, M. Aguilar et al., {\it First Result from the Alpha Magnetic Spectrometer on the International Space Station: Precision Measurement of the Positron Fraction in Primary cosmic-rays of 0.5-350 GeV, Phys. Rev. Lett.} {\bf 110} (2013) 141102.

\bibitem{Bergstrom2009} L. Bergstr\"om, J. Edsj\"o and G. Zaharijas, {\it Dark matter interpretation of recent electron and positron data, Phys. Rev. Lett.} {\bf 103} (2009) 031103 [arXiv:0905.0333].

\bibitem{Ibarra2014} A. Ibarra, A. S. Lamperstorfer and J. Silk, {\it Dark matter annihilations and decays after the AMS-02 positron measurements, Phys. Rev. D} {\bf 89} (2014) 063539 [arXiv:1309.2570].

\bibitem{Mauroa2014} M. Di Mauroa, F. Donato, N. Fornengo, R. Lineros and A. Vittino, {\it Interpretation of AMS-02 electrons and positrons data, J. Cosm. Astropart. Phys.} {\bf 04} (2014) 006 [arXiv:1402.0321].

%\bibitem{Giesen2015} G. Giesen et al., {\it AMS-02 antiprotons, at last! Secondary astrophysical component and immediate implications for Dark Matter, J. Cosm. Astropart. Phys.} {\bf 09} (2015) 023.

\bibitem{Jin2013} H. B. Jin, Y. L. Wu and Y. F. Zhou, {\it Implications of the first AMS-02 measurement for dark matter annihilation and decay, J. Cosm. Astropart. Phys.} {\bf 11} (2013) 026 [arXiv:1304.1997].

\bibitem{Yuan2013}  Q. Yuan et al., {\it Implications of the AMS-02 positron fraction in cosmic-rays, Astropart. Phys.} {\bf 60} 2015 1-12 [arXiv:1304.1482].

\bibitem{Ferrer2013} F. Ferrer and D. R. Hunter, {\it The impact of the phase-space density on the indirect detection of dark matter, J. Cosm. Astropart. Phys.} {\bf 09} (2013) 005 [arXiv:1306.6586].

\bibitem{Boddy2017} K. K. Boddy, J. Kumar, L. E. Strigari, {\it M. Y. Wang Sommerfeld-Enhanced $J$-factors For Dwarf Spheroidal Galaxies, Phys. Rev. D} {\bf 95} (2017) 123008 [arXiv:1702.00408].

\bibitem{Campbell2017} D. J. R. Campbell et al., {\it Knowing the unknowns: uncertainties in simple estimators of galactic dynamical masses, Mon. Not. Roy. Astron. Soc.} {\bf 469} (2017) 2 [arXiv:1603.04443].

\bibitem{Binney2008} J. Binney and S. Tremaine, {\it Galactic Dynamics: Second Edition,} Princeton University Press, Princeton U.S.A. (2008).

\bibitem{NFW1997} J. F. Navarro, C. S. Frenk and S. D. White, {\it A Universal density profile from hierarchical clustering, Astrophys. J.} {\bf 490} (1997) 493 [astro-ph/9611107].

\bibitem{Martinez2015} G. D. Martinez, {\it A Robust Determination of Milky Way Satellite Properties using hierarchical mass modelling Mon. Not. Roy. Astron. Soc.} {\bf 451} (2015) 2524 [arXiv:1309.2641].

\bibitem{Bonnivard2015} V. Bonnivard et al., {\it Dark matter annihilation and decay in dwarf spheroidal galaxies: the classical and ultrafaint dSphs, Mon. Not. Roy. Astron. Soc.} {\bf 453} (2015) 849 [arXiv:1504.02048].

\bibitem{Kuhlen2012} M. Kuhlen, M. Vogelsberger and R. Angulo, {\it Numerical Simulations of the Dark Universe: State of the Art and the Next Decade, Phys. Dark Univ.} {\bf 1} (2012) 50 [arXiv:1209.5745].

\bibitem{Diemand2011} J. Diemand and B. Moore, {\it The structure and evolution of cold dark matter halos, Adv. Sci. Lett.} {\bf 4} (2011) 297 [arXiv:0906.4340].

\bibitem{Blok2010} W. de Blok, {\it The Core-Cusp Problem, Adv. Astron.} {\bf 2010} (2010) 789293 [arXiv:0910.3538].

\bibitem{Spergel2000} D. N. Spergel and P. J. Steinhardt, {\it Observational evidence for self-interacting cold dark matter, Phys. Rev. Lett.} {\bf 84} (2000) 3760 [arXiv:astro-ph/9909386].

\bibitem{Vogelsberger2012} M. Vogelsberger, J. Zavala and A. Loeb, {\it Subhaloes in self-interacting galactic dark matter haloes, Mon. Not. R. Astron. Soc.} {\bf 423} (2012) 3740–3752.

\bibitem{Zavala2013} J. Zavala, M. Vogelsberger and M. G. Walker, {\it Constraining self-interacting dark matter with the Milky Way's dwarf spheroidals, Mon. Not. R. Astron. Soc. Lett.} {\bf 431} (2013) L20–L24.

\bibitem{Rocha2013} M. Rocha, et al., {\it Cosmological simulations with self-interacting dark matter $-$ I. Constant-density cores and substructure, Mon. Not. R. Astron. Soc.} {\bf 430} (2013) 81–104.

\bibitem{Peter2013} A. H. G. Peter, M. Rocha, J. S. Bullock and M. Kaplinghat, {\it Cosmological simulations with self-interacting dark matter $-$ II. Halo shapes versus observations, Mon. Not. R. Astron. Soc.} {\bf 430} (2013) 105–120.

\bibitem{Kaplinghat2016} M. Kaplinghat, S. Tulin and H. B. Yu, {\it Dark Matter Halos as Particle Colliders: Unified Solution to Small-Scale Structure Puzzles from Dwarfs to Clusters, Phys. Rev. Lett.} {\bf 116} (2016) 041302 [arXiv:1508.03339].

\bibitem{Tulin2013} S. Tulin, H. B. Yu and K. M. Zurek, {\it Resonant Dark Forces and Small-Scale Structure, Phys. Rev. Lett.} {\bf 110} (2013) 111301 [arXiv:1210.0900].

\bibitem{Kamada2017} A. Kamada, M. Kaplinghat, A. B. Pace and H. B. Yu, {\it Self-Interacting Dark Matter Can Explain Diverse Galactic Rotation Curves, Phys. Rev. Lett.} {\bf 119} (2017) 111102 [arXiv:1611.02716].

\bibitem{Lattanzi2009} M. Lattanzi, {\it Can the WIMP annihilation boost factor be boosted by the Sommerfeld enhancement?, Phys. Rev. D} {\bf 79} (2009) 083523 [arXiv:0812.0360].

\bibitem{Cassel2010} S. Cassel, {\it Sommerfeld factor for arbitrary partial wave processes, J. Phys. G} {\bf 37} (2010) 105009 [arXiv:0903.5307].

\bibitem{Slatyer2010} T. R. Slatyer, {\it The Sommerfeld enhancement for dark matter with an excited state, J. Cosmol. Astropart. Phys.} {\bf 02} (2010) 028 [arXiv:0910.5713].

\bibitem{lu2016-1} B. Q. Lu and H. S. Zong, {\it Limits on dark matter from AMS-02 antiproton and positron fraction data, Phys. Rev. D} {\bf 93} (2016) 103517 [arXiv:1510.04032].

\bibitem{Sameth2015} A. G. Sameth, S. M. Koushiappas and M. Walker, {\it DWARF GALAXY ANNIHILATION AND DECAY EMISSION PROFILES FOR DARK MATTER EXPERIMENTS Astrophys. J.} {\bf 801} (2015) 74
[arXiv:1408.0002].

\bibitem{Sanders2016} J. L. Sanders, N. W. Evans, A. G. Sameth and W. Dehnen, {\it Indirect Dark Matter Detection for Flattened Dwarf Galaxies Phys. Rev. D} {\bf 94} (2016) 063521 [arXiv:1604.05493].

\bibitem{Evans2016} N. W. Evans, J. L. Sanders and A. Geringer-Sameth, {\it Simple $J$-factors and D-Factors for Indirect Dark Matter Detection Phys. Rev. D} {\bf 93} (2016) 103512 [arXiv:1604.05599].

\bibitem{1FGL} Fermi-LAT collaboration, A. A. Abdo et al., {\it Fermi large area telescope first source catalog, Astrophys. J. Suppl.} {\bf 188} (2010) 405-436 [arXiv:1002.2280].

\bibitem{Berlin2014} A. Berlin and D. Hooper, {\it Stringent constraints on the dark matter annihilation cross section from subhalo searches with the Fermi Gamma-Ray Space Telescope, Phys. Rev. D} {\bf 89} (2014) 016014 [arXiv:1309.0525].

\bibitem{Bertoni2015} B. Bertoni, D. Hooper and T. Linden, {\it Examining The Fermi-LAT Third Source Catalog in search of dark matter subhalos, J. Cosmol. Astropart. Phys.} {\bf 12} (2015) 035 [arXiv:1504.02087].

\bibitem{Hooper2017} D. Hooper and S. J. Witte, {\it Gamma Rays From Dark Matter Subhalos Revisited: Refining the Predictions and Constraints, J. Cosmol. Astropart. Phys.} {\bf 04} (2017) 018 [arXiv:1610.07587].

\bibitem{3FGL} Fermi-LAT collaboration, F. Acero et al., {\it Fermi Large Area Telescope third source catalog, Astrophys. J. Suppl.} {\bf 218} (2015) 23 [arXiv:1501.02003].

\bibitem{VLII2008} J. Diemand et al., {\it Clumps and streams in the local dark matter distribution, Nature} {\bf 454} (2008) 735–738, [arXiv:0805.1244].

\bibitem{ELVIS2013} S. G. Kimmel, M. Boylan-Kolchin, J. Bullock and K. Lee, {\it ELVIS: Exploring the Local Volume in Simulations, Mon. Not. Roy. Astron. Soc.} {\bf 438} (2014) 2578–2596, [arXiv:1310.6746].

\bibitem{Gehrels1986} N. Gehrels, {\it Confidence limits for small numbers of events in astrophysical data, Astrophys. J.} {\bf 303} (1986) 336.

\bibitem{Fermi2015}  Fermi-LAT collaboration, M. Ackermann et al., {\it Searching for Dark Matter Annihilation from Milky Way Dwarf Spheroidal Galaxies with Six Years of Fermi Large Area Telescope Data, Phys. Rev. Lett.} {\bf 115} (2015) 231301.

\bibitem{online} {\it http://www-glast.stanford.edu/pub$\_$data/1048/.}

\bibitem{Fermi2014} Fermi-LAT collaboration, M. Ackermann et al., {\it Dark matter constraints from observations of 25 Milky Way satellite galaxies with the Fermi Large Area Telescope, Phys. Rev. D} {\bf 89} (2014) 042001 [arXiv:1310.0828].

\bibitem{Bringmann2017} T. Bringmann and P. Walia, {\it Strong Constraints on Self-Interacting Dark Matter with Light Mediators, Phys. Rev. Lett.} {\bf 118} (2017) 141802 [arXiv:1612.00845].

\bibitem{Blum2016} K. Blum, R. Sato and T. R. Slatyer, {\it Self-consistent calculation of the Sommerfeld enhancement, J. Cosmol. Astropart. Phys.} {\bf 6} (2016) 021 [arXiv:1603.01383].

\end{thebibliography}
\end{document}